\documentclass[nohyper,11pt,letterpaper]{JHEP3}
\usepackage[dvips]{epsfig}
\usepackage{amsmath}

\newcommand{\be}{\begin{equation}}
\newcommand{\ee}{\end{equation}}
\newcommand{\bea}{\begin{eqnarray}}
\newcommand{\eea}{\end{eqnarray}}
\newcommand{\ba}{\begin{eqnarray}}
\newcommand{\ea}{\end{eqnarray}}
\newcommand{\nn}{\nonumber \\}
\newcommand{\eqn}[1]{(\ref{#1})}
\newcommand{\beq}{\begin{equation}}
\newcommand{\eeq}{\end{equation}}
\newcommand{\beqa}{\begin{eqnarray}}
\newcommand{\eeqa}{\end{eqnarray}}
\newcommand{\beqar}{\begin{eqnarray*}}
\newcommand{\eeqar}{\end{eqnarray*}}

\newcommand{\reef}[1]{(\ref{#1})}

\newcommand{\eg}{{\it e.g.,}\ }
\newcommand{\ie}{{\it i.e.,}\ }

\newcommand{\dd}{\tilde{d}}

\def\nc {N_\mt{c}}
\def\nf {N_\mt{f}}

\def\t6 {T_\mt{D6}}

\newcommand{\te}{t_\mt{E}}
\newcommand{\td}{T_\mt{d}}
\newcommand{\tf}{T_\mt{fun}}

\newcommand{\mq}{M_\mt{q}}      
\newcommand{\qc}{\langle \tilde{\psi} \psi \rangle} 
\newcommand{\N}{{\cal N}} 
\newcommand{\mbar}{\bar{M}}
\newcommand{\rhomin}{{\rho_\mt{min}}}
\newcommand{\rhomax}{{\rho_\mt{max}}}

\newcommand{\ls}{\ell_\mt{s}}

\newcommand{\ids}{I_\mt{D7}}
\newcommand{\tids}{\tilde{I}_\mt{D7}}
\newcommand{\ide}{I_\mt{E}}
\newcommand{\tide}{\tilde{I}_\mt{E}}

\newcommand{\nb}{n_\mt{b}}
\newcommand{\nq}{n_\mt{q}}

\newcommand{\mt}[1]{\textrm{\tiny #1}}

\newcommand{\trho}{\varrho}  

\def\sac{\, , \,\,\,\,\,}

\newcommand{\lam}{\lambda}

\newcommand{\labell}[1]{\label{#1}} 

\newcommand{\pa}{\partial}

\newcommand{\ft}{\tilde{f}}

\newcommand{\tE}{E}
\newcommand{\D}{d} 

\newcommand{\ra}{\rightarrow}
\newcommand{\overlrarrow}[1]{\vbox{\ialign{##\cr\cr
                  \leftrightarrowfill\crcr\noalign{\kern-1pt\nointerlineskip}
                  $\hfil\displaystyle{#1}\hfil$\crcr}}}
\newcommand{\mub}{\mu_\mt{b}}

\title{Holographic phase transitions at finite baryon density}

\author{Shinpei Kobayashi,$^{a,b}$ David Mateos,$^c$ Shunji Matsuura,$^{a,d}$
Robert C. Myers,$^{a,b,e}$ and Rowan M. Thomson$^{a,b}$\\
$^a$ Perimeter Institute for Theoretical Physics,
Waterloo, Ontario N2L 2Y5, Canada \\
$^b$ Department of Physics and Astronomy, University of Waterloo,
Waterloo, Ontario\\
\ \  N2L 3G1, Canada \\
$^c$ Department of Physics, University of California, Santa Barbara, CA 93106-9530, USA\\
$^d$ Department of Physics, University of Tokyo,
7-3-1 Hongo, Bunkyoku, Tokyo\\
\ \ 113-0033, Japan\\
$^e$ Kavli Institute for Theoretical Physics, University of
California, Santa Barbara, CA\\
\ \ 93106-4030, USA\\
\\E-mail: \email{skobayashi@perimeterinstitute.ca,
dmateos@physics.ucsb.edu, smatsuura@perimeterinstitute.ca,
rmyers@perimeterinstitute.ca, rthomson@perimeterinstitute.ca }}

\abstract{We use holographic techniques to study $SU(\nc)$ super
Yang-Mills theory coupled to $\nf \ll \nc$ flavours of fundamental
matter at finite temperature and baryon density. We focus on four
dimensions, for which the dual description consists of $\nf$
D7-branes in the background of $\nc$ black D3-branes, but our
results apply in other dimensions as well. A non-zero chemical
potential $\mu_\mt{b}$ or baryon number density $\nb$ is introduced
via a nonvanishing worldvolume gauge field on the D7-branes.
Ref.~\cite{prl} identified a first order phase transition at zero
density associated with `melting' of the mesons. This extends to a
line of phase transitions for small $\nb$, which terminates at a
critical point at finite $\nb$. Investigation of the D7-branes'
thermodynamics reveals that $(\pa \mu_\mt{b} / \pa \nb)_T <0$ in a
small region of the phase diagram, indicating an instability. We
comment on a possible new phase which may appear in this region.}


\preprint{hep-th/0611099}

\begin{document}


\section{Introduction}

In strongly coupled, large-$\nc$ gauge theories with a gravity dual
\cite{juan,bigRev}, $\nf \ll \nc$ flavours of fundamental matter can be
described by $\nf$ D-brane probes \cite{flavour} in the appropriate
gravitational background. At sufficiently high temperatures, the
latter contains a black hole \cite{witten}. Working with $\nf \ll
\nc$ flavours ensures that the matter branes only make a small
perturbation to this background. Then much of the physics can be
studied in the probe approximation where the gravitational
backreaction of these branes is neglected.\footnote{The backreaction
can not be ignored in calculating the effect of the fundamental
matter on hydrodynamic transport coefficients such as the shear
viscosity \cite{visco}.}

\FIGURE{
 \includegraphics[width=0.95 \textwidth]{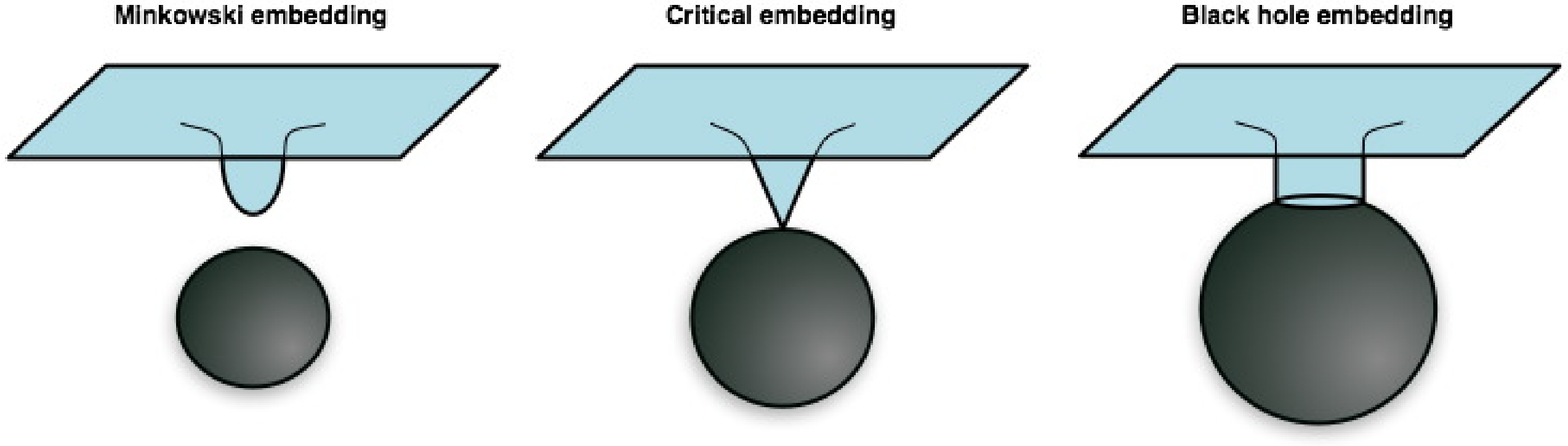}
\caption{Various possible D7-brane embeddings in the black D3-brane
geometry for zero baryon number density.  The temperature increases
from left to right. At finite $\nb$, the Minkowski (and critical)
embeddings are {\it not} allowed -- see discussion in the text.}
\label{embeddings2}}
This framework was recently used in \cite{prl,long,recent,recent8}
to study the thermal properties of $\nf$ flavours of fundamental
quarks (and scalars) in $SU(\nc)$ super Yang-Mills theories in
diverse dimensions.\footnote{Initial studies include
\cite{pions,toward}.} It was shown that a universal, first order
phase transition occurs at some critical temperature $\tf$. At low
temperatures, the branes sit outside the black hole in what was
dubbed a `Minkowski' embedding (see figure \ref{embeddings2}), and
stable meson bound states exist. In this phase the meson spectrum
exhibits a mass gap and is discrete. Above some critical temperature
$\tf$ the branes fall through the horizon in what were dubbed `black
hole' embeddings. In this phase the meson spectrum is gapless and
continuous. This large-$\nc$, strong coupling phase transition is
therefore associated with the melting of the mesons. In theories
that undergo a confinement/deconfinement phase transition at some
temperature $\td < \tf$, mesonic states thus remain bound in the
deconfined phase for the range of temperatures $\td < T < \tf$.

This physics is in qualitative agreement with that of QCD, in which
$s \bar{s}$ and $c\bar{c}$ states, for example, seem to survive the
deconfinement phase transition at $\td \simeq 175$ MeV -- see
\cite{long} for a more detailed discussion. It is thus interesting
to ask how this physics is modified at finite baryon density. In the
presence of $\nf$ flavours of  equal mass, the gauge theory
possesses a global $U(\nf) \simeq SU(\nf) \times U(1)_\mt{q}$
symmetry. The $U(1)_\mt{q}$ charge counts the net number of quarks,
\ie the number of baryons times $\nc$ -- see appendix \ref{holo} for
details. In the gravity description, this global symmetry
corresponds to the $U(\nf)$ gauge symmetry on the worldvolume of the
$\nf$ D-brane probes. The conserved currents associated to the
$U(\nf)$ symmetry of the gauge theory are dual to the gauge fields
on the D-branes. Thus, the introduction of a chemical potential
$\mub$ or a non-zero density $\nb$ for the baryon number in the
gauge theory corresponds to turning on the diagonal $U(1) \subset
U(\nf)$ gauge field on the D-branes.\footnote{This should not be
confused with the chemical potential for R-charge (as considered in,
\eg \cite{charge,charge2}) which is dual to internal angular
momentum on the $S^5$ in the gravity description.} 
\FIGURE{
 \includegraphics[width=0.45 \textwidth]{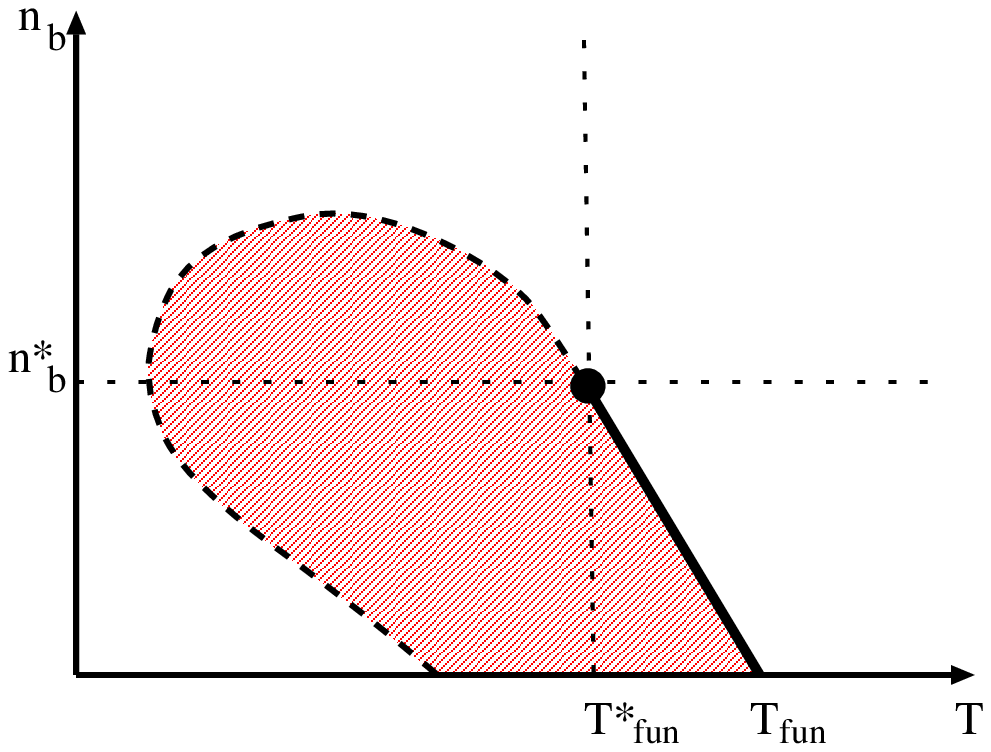}
\caption{Phase diagram: Baryon number $n_b$ versus temperature $T$.
The line of first order phase transitions ends with a critical point
at $(T_\mt{fun}^*,\nb^*)$. The phase which we study is intrinsically
unstable in the shaded (red) region. This plot shows only a small
portion of the full phase diagram near the critical point. The
origin of the axes above corresponds to
$(\nb,T)=(0,0.986\,T_\mt{fun})$.} \label{phase-diagram}}

In this paper we study the gauge theory at constant baryon number
density $\nb$.
We find that, for any finite value of the baryon number
density, the Minkowski embeddings, \ie those embeddings where the
probe brane closes off above the horizon, are physically
inconsistent. Hence at finite $\nb$, we focus our study on black
hole embeddings. Despite this difference with the $\nb=0$ case, the
first order phase transition found there continues to exist here for
small enough a baryon number density. In this case, however, the
transition is between two black hole embeddings. For a large enough
baryon number density, there is no phase transition as a function of
the temperature. The phase transition ceases to exist at a critical
value $\nb^*$. These results are summarised in fig.
\ref{phase-diagram}. This phase diagram also shows a shaded region
where the black hole embeddings are found to be thermodynamically
unstable. While the boundary of this region shown in the diagram is
qualitative, we have found that the unstable region has a limited
extent to the left of the line of first order phase transitions.
Hence the system must find a new stable phase, at least, in this
small region -- see section \ref{thermal}.

We focus on four-dimensional  ${\cal N}=4$ super Yang-Mills coupled
to fundamental matter, whose dual description consists of $\nf$
D7-branes in the background of $\nc$ black D3-branes, but our
results hold in other dimensions. Investigations of other
holographic systems with a chemical potential have appeared
previously in \cite{chemical,chemical8}. An overview of the paper is
as follows:  In section \ref{Geom}, we solve for the embedding of
the D7-branes in the black D3-brane geometry. Our discussion
includes a brief review of the black hole background, the equations
of motion determining the embedding, and a careful analysis of the
required boundary conditions. In this section, we also discuss the
effect of finite $\nb$ on the critical solution and self-similar
scaling found at $\nb=0$ \cite{prl,long}. Finally we present the
results of numerically solving for the embeddings at various values
of the baryon number density. Section \ref{thermal} examines the
thermal properties of the D7-branes, including their stability or
lack thereof. Section \ref{discuss} concludes with a discussion of
results. Appendix \ref{holo} presents some details of the
holographic dictionary relating the worldvolume fields describing
the D7-brane embeddings to their dual operators in the gauge theory.

\section{Holographic framework}
\label{Geom}

\subsection{Black D3-branes}

As first proposed in \cite{juan}, ${\cal N}=4$ super-Yang-Mills
(SYM) with gauge group $SU(\nc)$ is holographically dual to type IIB
string theory on $AdS_5\times S^5$ with $\nc$ units of RR five-form
flux. The dictionary relating the two sides of the duality equates
$g_s=g^2_\mt{YM}/2\pi$ and $L^4/\ls^4= 2g^2_\mt{YM}\nc\equiv2\lam$,
where $L$ is the AdS curvature scale -- for a review, see
\cite{bigRev}. In the limit of large $\nc$ and large $\lam$, the
string side of the duality reduces to (weakly coupled) classical
gravity. At a finite temperature, a black hole appears in the
supergravity background \cite{witten}. Following \cite{prl,long},
the black hole metric may be written as\footnote{This metric is
related to the standard presentation with the coordinate
transformation $\varrho^2 = u^2 + \sqrt{u^4-u_0^4}$.}
\beq ds^2 = \frac{1}{2} \left(\frac{\varrho}{L}\right)^2
\left[-{f^2\over \tilde f}dt^2 + \tilde{f} dx^2_3 \right]
 + \frac{L^2}{\varrho^2}\left[ d\varrho^2 +\varrho^2 d\Omega_5^2
  \right] \,,
  \labell{D3geom}
\eeq
where
\be f(\trho)= 1- \frac{u_0^4}{\trho^4} \sac
\tilde{f}(\trho)=1+\frac{u_0^4}{\trho^4} \,. \ee
The gauge theory temperature is then equivalent to the Hawking
temperature of the black hole horizon, determined as usual by the
surface gravity $T=\kappa/2\pi$. Alternatively, the latter
temperature may be determined by demanding regularity of the
Euclidean section obtained through the Wick rotation $t \rightarrow
i\te$. Then $\te$ must be periodically identified with a period
$\beta$, where
\be \frac{1}{\beta}=T = \frac{u_0}{\pi L^2}  \,. \labell{beta} \ee
This holographic framework allows the thermal behaviour of the
strongly coupled gauge theory to be further studied with standard
semiclassical gravity techniques \cite{hawk}. In particular, the
entropy density can be calculated as the geometric
Hawking-Bekenstein entropy of the horizon
\cite{witten,3/4}\footnote{We divide out by the (formally infinite)
three-dimensional volume $V_x$ of the Minkowski space in which the
gauge theory is formulated to yield the (finite) entropy density
\reef{entropy}.}
\be
S = \frac{\cal A}{4G\,V_x}=\frac{\pi^6}{4G}\,\frac{L^8}{\beta^3}
= \frac{\pi^2}{2} \nc^2\,T^3 \,, \labell{entropy} \ee
where we have used $16\pi G=(2\pi)^7 \ell_s^8\, g_s^2$. The
parametric dependence $S \propto\nc^2$ reflects the fact that the
gauge theory is deconfined. Remarkably, this strong coupling result
differs from that calculated at weak coupling by merely a factor of
$3/4$ \cite{3/4}.

\subsection{D7-brane embeddings}
\label{22}

One feature of the ${\cal N}=4$ SYM theory appearing in the duality
above is that all of the fields transform in the adjoint
representation of the $SU(\nc)$ gauge group. Fields transforming in
the fundamental representation can be included by introducing an
additional set of D-branes on the string theory side of the duality.
Following \cite{flavour}, we consider the decoupling limit of the
intersection of $\nc$ D3-branes and $\nf$ D7-branes as described by
the following array:
\begin{equation}
\begin{array}{ccccccccccc}
   & 0 & 1 & 2 & 3 & 4& 5 & 6 & 7 & 8 & 9\\
\mbox{D3:} & \times & \times & \times & \times & & &  &  & & \\
\mbox{D7:} & \times & \times & \times & \times & \times  & \times
& \times & \times &  &   \\
\end{array}\labell{D3D7}
\end{equation}
The resulting dual gauge theory is $\mathcal{N}=4$ super-Yang-Mills
coupled to $\nf$ $\mathcal{N}=2$ fundamental hypermultiplets
\cite{flavour} at temperature $T$ in $3+1$ dimensions. Assuming
$\nf\ll\nc$, the decoupling limit leads to $\nf$ probe D7-branes in
the previous background \reef{D3geom}, with the intersection being
parametrised by the coordinates $\{ t, x^i \}$. Since the D7-branes
span the 4567-directions, it is useful to introduce spherical
coordinates $\{ r, \Omega_3\}$ in this space and polar coordinates
$\{R, \phi\}$ in the 89-directions. Denoting by $\theta$ the angle
between these two spaces we then have:
\beq
\trho^2=r^2+R^2 \sac r=\trho\sin\theta \sac
R=\trho\cos\theta \,,
\labell{coord2}
\eeq
and
\beqa
d\trho^2+\trho^2d\Omega^2_5&=& d\trho^2+\trho^2\left(d\theta^2+\sin^2\theta
\, d\Omega^2_3+ \cos^2\theta \, d\phi^2\right) \labell{met2} \\
&=& dr^2 + r^2d \Omega^2_3 + dR^2+R^2d\phi^2 \,.
\labell{met1}
\eeqa

The analysis is simplified by taking $\chi=\cos\theta$ to describe
the embedding of the D7-branes.  Translational symmetry in the
0123-space and rotational symmetry in the 4567-directions motivate
us to take $\chi = \chi(\varrho)$. The induced metric on the
D7-branes is then:
\beq ds^2 = \frac{1}{2} \left(\frac{ \trho}{L}\right)^2
\left[-{f^2\over \tilde f}dt^2 + \tilde{f} dx^2_3 \right] +
\frac{L^2}{\trho^2}\left[ \frac{1-\chi^2 +
\trho^2(\partial_{\trho}\chi)^2}{1-\chi^2} \right] d\trho^2 +L^2
(1-\chi^2) d\Omega_3^2 \,. \labell{D3geom-induced} \eeq
We also introduce a $U(1)$ gauge field on the worldvolume of the
D7-branes. As we discuss in detail in appendix \ref{holo}, in order
to study the gauge theory at finite chemical potential or baryon
number density, it suffices to turn on the time component of the
gauge field, $A_t$. Again, symmetry considerations lead us to take
the ansatz $A_t = A_t(\varrho)$. The action of the D7-branes then
becomes:
\beq \ids=-\nf T_\mt{D7} \int d^8 \sigma \frac{\trho^3}{4} f
\tilde{f} (1-\chi^2) \sqrt{1-\chi^2+\trho^2 (\partial_\trho
\chi)^2-2(2 \pi \ls^2)^2 \frac{\tilde{f}}{f^2}(1-\chi^2) F_{\trho
t}^2} \,, \labell{action} \eeq
where $F_{\varrho t}=\partial_\varrho A_t$ is a radial electric field.

The equation of motion for $A_t$ (Gauss' law) gives
\beq
\partial_\trho \left[  \frac{\trho^3}{2}\frac{\tilde{f}^2}{f}
\frac{(1-\chi^2)^2 \partial_{\trho}A_t}{\sqrt{1-\chi^2+\trho^2
(\partial_\trho \chi)^2-2 (2 \pi \ls^2)^2 \frac{\tilde{f}}{f^2}
(1-\chi^2) (\partial_\trho A_t)^2}}\right] =0 \, . \labell{gauss}
\eeq
In the limit that $\trho \to \infty$, this equation reduces to
$\partial_\trho ( \trho^3\partial_{\trho}A_t ) \simeq0$ and so the
asymptotic solution approaches
\beq A_t \simeq \mu - \frac{a}{\trho^2} + \cdots\ .
\labell{asymptA_t} \eeq
The constants $\mu$ and $a$ are (proportional to) the chemical
potential for and the vacuum expectation value of the baryon number
density, respectively (see appendix \ref{holo}). The equation of
motion \reef{gauss} clearly indicates that there is a constant of
motion, which we write as
\beq d\equiv \nf T_{D7} (2 \pi \ls^2)^2
\frac{\trho^3}{2}\frac{\tilde{f}^2}{f} \frac{(1-\chi^2)^2
\partial_{\trho}A_t}{\sqrt{1-\chi^2+\trho^2(\partial_\trho \chi)^2
-2 (2 \pi \ls^2)^2 \frac{\tilde{f}}{f^2} (1-\chi^2) (\partial_\trho
A_t)^2}} \,. \labell{disp} \eeq
With this normalization, this constant is precisely the electric
displacement, $d= \delta \ids/\delta F_{\trho t}$. Taking the
large-$\trho$ limit of eq.~\reef{disp} with the asymptotic form
\reef{asymptA_t}, we find:
\beq d=\nf T_\mt{D7}(2 \pi \ls^2)^2\, a. \labell{d} \eeq

Now one could proceed to derive the equation of motion for the
D7-brane profile $\chi(\trho)$ from the action \reef{action} and
then use eq.~\reef{disp} to eliminate $A_t$ in favor of the constant
$d$. Instead, we first construct the Legendre transform of
eq.~\reef{action} with respect to $d$ to eliminate $A_t$ directly at
the level of the action. The result is:
\beqa
 \tids&=&\ids-\int d^8\, \sigma F_{\trho t}\,\frac{\delta I}{\delta
F_{\trho t}}\labell{actionp}\\
&=& -\nf T_\mt{D7} \int d^8 \sigma\, \frac{\trho^3}{4} f \tilde{f}
(1-\chi^2) \sqrt{1-\chi^2+\trho^2 (\partial_\trho
\chi)^2}\,\left[1+\frac{8\,d^2}{(2\pi\ls^2\nf T_\mt{D7})^2
\trho^6\tilde{f}^3(1-\chi^2)^3}\right]^{1/2}\,.\nonumber
 \eeqa
The gauge field equations resulting from this Legendre transform are
simply $\pa_\trho d= \delta\tids/\delta A_t$ and $\pa_\trho A_t=
-\delta\tids/\delta d$. The first of these reproduces the fact that
$d$ is a fixed constant and we will return to the second one below.

Before deriving the equation of motion for the D7-brane profile
$\chi(\trho)$, it is convenient to introduce dimensionless
quantities:
\beq \rho=\frac{\trho}{u_0}\, , \qquad \tilde{d} = \frac{d}{2 \pi
\ls^2 u_0^3 \nf T_\mt{D7}}\, . \labell{redef} \eeq
The $\chi$ equation from eq.~\reef{actionp} can then be written as
\beqa
&& \partial_\rho\left[\frac{\rho^5 f \tilde{f} (1-\chi^2)
\dot{\chi}}{\sqrt{1-\chi^2+\rho^2\dot{\chi}^2}} \sqrt{1 +
\frac{8 \tilde{d}^2}{\rho^6 \tilde{f}^3 (1-\chi^2)^3}}\right]\labell{chiEom} \\
&& =- \frac{\rho^3 f \tilde{f} \chi }{\sqrt{1-\chi^2+\rho^2\dot{\chi}^2}}
\sqrt{1 +\frac{8 \tilde{d}^2}{\rho^6 \tilde{f}^3 (1-\chi^2)^3}}
\left[3 (1-\chi^2) +2 \rho^2 \dot{\chi}^2 -24 \tilde{d}^2
\frac{1-\chi^2+\rho^2\dot{\chi}^2}{\rho^6 \tilde{f}^3 (1-\chi^2)^3+8 \tilde{d}^2}
\right]\,, \nonumber
\eeqa
where the dot denotes derivatives with respect to $\rho$, \ie
$\dot{\chi} = \partial _\rho \chi$. With $\rho \to \infty$, this
equation becomes at leading order: $\partial_\rho (\rho^5
\dot{\chi})\simeq-3\rho^3\,\chi$. Hence asymptotically the profile
behaves as
\beq
\chi = \frac{m}{\rho}+\frac{c}{\rho^3} + \cdots \,, \labell{asymptD7}
\eeq
where the dimensionless constants $m$ and $c$ are proportional to
the quark mass and condensate, respectively \cite{prl,long}. The
precise relations are given in appendix \ref{holo}.

Returning to the gauge field, we begin by introducing a convenient
dimensionless potential and chemical potential:
\beq \tilde{A}_t=\frac{2 \pi \ls^2}{u_0}\, A_t \sac
 \tilde{\mu} = \frac{2\pi \ls^2}{u_0}\,\mu \,. \labell{connect}\eeq
Then as described above, \reef{actionp} yields the following
equation
\beq \pa_\rho\tilde{A}_t = 2 \tilde{d}
\frac{f^2 \sqrt{1-\chi^2+\rho^2 \dot{\chi}^2}}
{\sqrt{\tilde{f}(1-\chi^2) [\rho^6 \tilde{f}^3 (1-\chi^2)^3+8 \tilde{d}^2]}}\,. \eeq
Integrating yields the potential difference between two radii,
\beq \tilde{A}_t(\rho)-\tilde{A}_t(\rho_0) = 2\tilde{d}
\int_{\rho_0}^\rho d\rho
\frac{f\sqrt{1-\chi^2+\rho^2\dot{\chi}^2}}{\sqrt{\tilde{f}
(1-\chi^2)[\rho^6 \tilde{f}^3 (1-\chi^2)^3 +8 \tilde{d}^2]}}\,.
\labell{gaugeField} \eeq
We will see below that all embeddings of interest extend down to the
horizon at $\rho=1$, so $\rho_0=1$ provides a convenient reference
point. Further we set $\tilde{A}_t(\rho=1)=0$ by the following
argument: The event horizon of the background \reef{D3geom} can be
characterized as a Killing horizon, which implies that it contains
the bifurcation surface where the Killing vector $\pa_t$ vanishes
\cite{kill}. If the potential $\tilde{A}$ as a one-form is to be
well defined, then $\tilde{A}_t$ must vanish there. Hence we can use
\reef{gaugeField} to calculate the chemical potential, \ie
$\tilde{A}_t(\infty)$, as
\beq \tilde{\mu}  = 2\tilde{d} \int_1^\infty d\rho\,
\frac{f\sqrt{1-\chi^2+\rho^2\dot{\chi}^2}}
{\sqrt{\tilde{f}(1-\chi^2)[\rho^6 \tilde{f}^3 (1-\chi^2)^3 +8
\tilde{d}^2]}}\,.\labell{moo} \eeq

\subsection{Near-horizon embeddings}
\label{new2}

An important role in \cite{prl,long} was played by the analysis of
the probe brane embeddings in the near-horizon region of the
geometry \eqn{D3geom}. In this section we will see how this analysis
is affected by the presence of the electric field on the D7-branes.
In fact we will generalize the analysis to consider probe Dq-branes
in a black Dp-brane background, along the lines of \cite{prl}. These
calculations will lead to two main conclusions. The first one is
that smooth Minkowski embeddings are unphysical for any non-zero
baryon density. The second one is that we expect the first order
phase transition found in \cite{prl,long} to persist for small
values of the baryon density, but to disappear for sufficiently
large densities.

In order to focus on the near-horizon region, we set \be \trho = u_0
+ \frac{L}{u_0} z \sac \theta = \frac{y}{L} \,, \ee and expand the
metric \eqn{D3geom} to lowest order in $z,y$. This yields Rindler
space together with some spectator directions which we omit since
they will play no role in the following:
\be ds^2 = - (2\pi T)^2 z^2 dt^2 + dz^2 + dy^2 + y^2 d\Omega_n^2 +
\cdots \,. \label{nearG}\ee
We recall that $T = u_0 /\pi L^2$. In \eqn{nearG} we have introduced
an integer $n$ equal to the dimension of the internal sphere wrapped
by the probe Dq-branes. For the D3/D7 system $n=3$, but as stated
above our analysis in this section will apply to more general Dp/Dq
systems, for which possibly $n\neq 3$; for example, $n=2$ for the
D4/D6 system of \cite{toward}. The horizon is of course at $z=0$.
The coordinates $z$ and $y$ are the near-horizon analogues of the
global coordinates $R$ and $r$ in \eqn{met1}, respectively.

In order to describe the embedding of the Dq-branes, we choose the
static gauge for all their coordinates except the radial coordinate
on the brane, which we denote as $\sigma$. The Dq-brane embedding
may then be described parametrically as: $z=z(\sigma)$,
$y=y(\sigma)$. We modify the analysis of \cite{prl} by adding a
radial electric field $E\equiv \ell_s^2 \dot{A_t} / T$, where the
dot denotes differentiation with respect to $\sigma$.  For
simplicity, in this section we will ignore the overall normalisation
of the Dq-branes action and take $I_\mt{D7} \propto \int d\sigma
{\cal L}$, where
\be {\cal L}  =  - y^n \sqrt{z^2 (\dot{z}^2 + \dot{y}^2) - E^2}
\,.
\ee
This action is homogeneous of degree $2+n$ under the rescaling
\be z \ra \alpha z \sac y \ra \alpha y \sac E \ra \alpha^2 E \,,
\labell{scal1} \ee
which means that the equations of motion will be invariant under
such a transformation. Recall that as first described in
\cite{frolov}, this scaling symmetry was a key ingredient for
self-similarity of the brane embeddings in \cite{prl,long}. However,
in the present case with $E\ne0$, the symmetry does not act within
the family of embedding solutions with a fixed electric field (or
rather fixed $d$ -- see eq.~\reef{scal2} below). Hence we can not
expect to find exactly the same self-similar behaviour for branes
supporting a fixed chemical potential or baryon density. However, we
argue below that the embeddings should behave in approximately the
same way at least where the gauge field is a small perturbation on
the Dq-brane.

As in the previous subsection, it is convenient to work with the
electric displacement
\be d = \frac{\pa {\cal L}}{\pa \tE} = \frac{y^n \tE}{\sqrt{ z^2
(\dot{z}^2 + \dot{y}^2) - \tE^2}} \,, \labell{displace} \ee
which is constant by virtue of Gauss' law. This is the near-horizon
analogue of the quantity with the same name introduced in the
previous subsection.\footnote{Note, however, that they differ in
their normalisation.} Note that under the scaling \reef{scal1} $\D$
transforms as
\be \D\ra\alpha^n\D\ .\label{scal2}\ee
Inverting the relation \reef{displace} above, one finds
\be \tE^2 = \frac{\D^2 z^2 (\dot{z}^2+\dot{y}^2) }{\D^2 + y^{2n}}
\,. \label{E} \ee
It is also useful to note the relation
\be \sqrt{z^2 (\dot{z}^2 + \dot{y}^2) - \tE^2} = y^{n} z\left[
\frac{\dot{z}^2 +\dot{y}^2 }{\D^2 + y^{2n}}\right]^{1/2} \,. \ee
To eliminate $\tE$ in favour of $\D$ and obtain a functional for
$y(\sigma)$ and $z(\sigma)$, we perform a Legendre transformation by
defining
\be \tilde{{\cal L}} =   {\cal L} - E \D  = - z\, \sqrt{\dot{z}^2 +
\dot{y}^2} \sqrt{\D^2 + y^{2n}} \,, \ee
in analogy with \eqn{actionp}. It is easily verified that the
equations of motion obtained from $\tilde{{\cal L}}$ are the same as
those obtained by first varying ${\cal L}$ and then using
eq.~\eqn{E} to eliminate $\tE$.

We can conclude from eq.~\eqn{E} that Minkowski embeddings which
close off smoothly at the $y$-axis, such as those considered in
\cite{prl,long}, are unphysical if $d \neq 0$. These embeddings are
most appropriately described in the gauge $y=\sigma$, and they are
characterised by the condition that the brane reaches $y=0$ at some
finite $z=z_0>0$. For the brane geometry to be smooth there, we must
impose the boundary condition $\dot{z}(0)=0$. Eq.~\eqn{E} then
yields $E^2= z_0^2$ at $y=0$. Now even though $E$ remains finite,
the tensor field $E dy \wedge dt$ is ill-defined at the origin and
so one should conclude that these configurations are singular. This
singularity is made clearer by considering the electric displacement
$d$ which also remains constant at the origin. However, one should
note that as defined in eq.~\reef{displace} $d$ is actually a tensor
density and so the norm of the associated tensor field is $\left|
\frac{d}{\sqrt{-g}} \frac{\pa}{\pa y} \frac{\pa}{\pa t} \right|^2
\sim d^2/y^{2n}$, which clearly diverges at the origin. The
physical reason why Minkowski embeddings are inconsistent is, of
course, that the radial electric field lines have nowhere to end if
the brane closes off above the horizon. This makes it clear that,
although we have obtained this result in the near-horizon
approximation, the same conclusion follows from an analysis in the
full geometry \eqn{D3geom}.

For D-branes, an electric field on the worldvolume can also be
associated with fundamental strings `dissolved' into the the
D7-brane \cite{bound} -- see also the discussion around
eq.~\reef{equiv}. Hence the above statement that the electric field
lines have nowhere to end can also be viewed as the fact that the
strings have nowhere to end if the brane closes off. However, rather
than simply viewing the Minkowski embeddings as unphysical, this
point of view lends itself to the interpretation that these
embeddings by themselves are incomplete. That is, one could imagine
constructing a physical configuration by attaching a bundle of
fundamental strings to the brane at $y=0$ and letting these stretch
down to the horizon. The strings resolve the singularity in the
electric field since they act as point charges which are the source
of this field. However, in such a configuration, the strings and the
brane must satisfy a force balance equation at the point where they
are connected. It is clear that if the brane closes off smoothly
with $\dot{z}(0)=0$, then they can not exert any vertical force in
the $z$ direction to balance the tension of the strings and so this
can not be an equilibrium configuration. One might then consider
`cuspy' configurations which close off with a finite $\dot{z}(0)$
but still at some $z=z_0>0$. In this case, the branes exert a
vertical force and so one must examine the configuration in more
detail to determine if the two forces can precisely balance. This
analysis requires a more careful treatment of the normalisation of
the brane action and the fields than we have presented here. Hence
we defer the detailed calculations to the next subsection where we
will examine the D7-branes in more detail. However, let us  state
the conclusion here: no Minkowski embeddings can achieve an
equilibrium for any (finite) value of $\dot{z}(0)$. Therefore we
discard Minkowski embeddings for the rest of our analysis in the
following.

Hence we now turn to consider black hole embeddings which intersect
the horizon. Since these reach the horizon $z=0$ at some $y=y_0$
they are conveniently described in the gauge $z=\sigma$. The
appropriate boundary condition in this case is then $\dot{y}(0)=0$,
and the equation that follows from $\tilde{\cal L}$ is
\be (y^{2n} + \D^2)
\left[ z y \ddot{y} + (1 + \dot{y}^2) y \dot{y} \right] - y^{2n} (1
+ \dot{y}^2) n z =0 \,. \label{eom} \ee
In view of this equation it is clear that we should expect two
qualitatively different behaviours for  solutions with $y_0^n \gg d$
and $y_0^n \ll d$. In the first case, it is easy to see that $y^n
\gg d$ all along the solution, and so we effectively recover the
equations of motion for $d=0$ studied in \cite{prl,long}, and
therefore oscillatory behaviour around a critical solution for large
$y$:
\be y \simeq
\sqrt{n} \, z + \xi \sac \xi = \frac{T^{-1}}{(T z)^{\frac{n}{2}}}
\left[ a \, \sin ( \alpha \log T z) + b \, \cos ( \alpha \log T z)
\right] \,, \ee
where $a,b$ are determined by $y_0$. As shown in \cite{prl}, this
oscillatory behaviour eventually leads to the property that the
quark condensate is multi-valued as a function of the quark mass,
and hence to a first order phase transition (see figure
\ref{cVsTd10m6} and the discussion in the next subsection). We thus
expect a similar transition if $y_0^n \gg d$.

Incidentally, note that, unlike in the case $d=0$, here the
`critical solution' $y=\sqrt{n} \, z$ is not an exact solution of
eq. \eqn{eom} but only an approximate solution for large $y$. In
particular, there is no exact solution of the form $y \propto z$
that just touches the horizon except the $y=0$ solution. Note also
that for black hole embeddings eq. \eqn{E} gives $E \sim z$ as $z
\ra 0$, leading to a well defined tensor field at the horizon $z=0$.

We now turn to the case $y_0^n \ll d$, for which the equation of
motion \reef{eom} reduces to
\be z \ddot{y} + (1 + \dot{y}^2)
\dot{y}  \simeq 0 \,, \label{eomsmall} \ee
whose exact solution is
\bea
\dot{y}&=&{z_1\over\sqrt{z^2- z_1^2}}\ ,\label{h1}\\
y&=&y_0+z_1\,\log\left({z +\sqrt{{z^2 -z_1^2}}}\right)c\,,
\label{h2}
\eea
where $y_0$ and $z_1$ are integration constants.
Recall that the boundary conditions should be $y(z=0)=y_0$ and
$\dot{y}(z=0)=0$. It is impossible to satisfy these
conditions with the logarithm in eq.~\reef{h2}. It is also clearly
seen in eq.~\reef{h1} that the general solution is problematic (at
$z=z_1$) unless $z_1=0$. Hence the only physical solution in this
regime is precisely the constant solution: $y=y_0$.

Further, we note that the embedding starts very near the horizon
with $y=y_0$ where $y_0^n\ll \D$ and so we ask how it makes a
transition to some more interesting profile of the full equation
\reef{eom} far from the horizon. The point is that the term $n
y^{2n} z$ will eventually grow large and require $y$ to deviate from
a constant. Quantitatively, one finds that the transition occurs for
$z\sim y_0\, (d/y_0^n)$ where the leading solution has the form
\be y = y_0 + {n\over 4} \left({y_0^n\over d}\right)^2 {z^2\over
y_0} + \cdots \,. \label{near1}
\ee
Hence we see the $O(z^2)$ correction to the constant embedding is
enormously suppressed in this regime $y_0^n\ll \D$. Note that at
$z\sim y_0\, (d/y_0^n)$, the second term is comparable to the first
and so the Taylor series is breaking down. However, at this point,
we still have $y^n\ll d$ and $\dot{y}\ll 1$. In summary, the
solution in this regime is a long spike that emanates from the
horizon almost vertically, resembling a bundle of strings.

The analysis above thus leads to the following physical picture. If
$d$ is small enough, then there is a set of embeddings in the
near-horizon region for which $y_0^n \gg d$, whose physics is
similar to that of the $d=0$ case. In particular, we expect a first
order phase transition to occur as a function of the temperature. As
$d$ increases, the region where the condition  $y_0^n \gg d$ holds
gets pushed outside the regime in which the near-horizon analysis is
applicable, suggesting that the phase transition as a function of
temperature should cease to exist for sufficiently large $d$. This
is precisely what the phase diagram in figure \ref{phase-diagram}
confirms. In contrast, the condition $y_0^n \ll d$ can always be met
in the near-horizon region, indicating that solutions for which the
part of the brane near the horizon behaves as a narrow cylinder of
almost constant size, resembling a bundle of strings, exist for all
values of $d$. This is also confirmed by our numerical analysis in
the full geometry (as illustrated in figure \ref{embeddings}), since
such type of embeddings can always be realised, for any fixed $d$,
by increasing the quark mass (or equivalently by decreasing the
temperature). In the next subsection we analyse some properties of
these embeddings more closely.

\subsection{Strings from branes}
\label{new}

The near-horizon analysis above revealed the existence of solutions
for which the brane resembles a long narrow cylinder that emanates
from the horizon. One's intuition is that this spike represents a
bundle of strings stretching between the asymptotic brane and the
black hole. Examples in which fundamental strings attached to a
D-brane are described as an electrically charged spike solution of
the DBI action are well known in flat space \cite{death}, in AdS
space \cite{baryon-vertex} and in other brane backgrounds
\cite{junctions}. Here we would like to formalise this intuition by
investigating the core region of our D7-brane embeddings in more
detail. This analysis allows us to investigate the boundary
conditions for the Minkowski-like embeddings in detail.

We begin by rewriting the Legendre-transformed action \reef{actionp} as
\beq \tids=-\frac{T_\mt{D7}}{\sqrt{2}} \int d^8 \sigma
\frac{f}{\ft^{1/2}}\sqrt{1+\frac{\trho^2 (\partial_\trho
\chi)^2}{1-\chi^2}}\,\left[\frac{d^2}{(2\pi\ls^2T_\mt{D7})^2}+
\frac{\nf^2}{8}\trho^6\tilde{f}^3(1-\chi^2)^3\right]^{1/2} \,.
\labell{actionp2} \eeq
Now recall that $\chi=\cos\theta$ -- see eq.~\reef{met2} -- and
consider the last factor in the integrand. If the embedding is very
near the axis, \ie $\chi\simeq 1$, then the second contribution in
this factor can be neglected and eq.~\reef{actionp2} becomes
\bea \tids &\simeq& - \nq V_x \frac{1}{2\pi \ls^2} \int dt\,d\trho\,
\frac{f}{(2\ft)^{1/2}}\sqrt{1+\trho^2 (\partial_\trho \theta)^2} \nn
&=&- \nq V_x \frac{1}{2\pi \ls^2} \int dt\,d\trho\,
\sqrt{-g_{tt}\left(g_{\trho\trho}+g_{\theta\theta}(\partial_\trho
\theta)^2\right)} \,,
\labell{actionp3}
\eea
where we have used the relation \eqn{ad} between $d$ and the density
of strings $\nq$. We recognize the result above as the Nambu-Goto
action for a bundle of fundamental strings stretching in the $\trho$
direction but free to bend away from $\theta=0$  on the $S^5$. It is
interesting to note that the term that was dropped provides
precisely the measure factor associated with the $x^i$ and $S^3$
directions in the limit where the $d$ term vanishes (or is small).
In this sense then, the D7-brane forgets about its extent in those
directions.

Let us consider the boundary conditions for the configurations which
reach the axis $\theta=0$ at some finite $\trho$, \ie for
Minkowski-like embeddings. These embeddings would in general have a
cusp if $\pa_\trho\theta$ remains finite at $\theta=0$ (a smooth
embedding would correspond to $\pa_\trho\theta\ra \infty$). As
discussed in the previous subsection, to produce a potentially
physical configuration, we would attach a bundle fundamental strings
to the tip of the brane (with precisely the density $\nq$). However,
to produce a consistent static configuration, there must be a
balance between the forces exerted by these external strings and the
brane along the $\trho$-direction. The effective tension of the
branes can be evaluated in many ways, but here we consider the
calculation:
\beq T_{\trho\trho}=\frac{2}{\sqrt{-g}}\,\frac{\delta\tids}{\delta
g^{\trho\trho}} \simeq\nq V_x\, \frac{1}{2\pi \ls^2}
\frac{g_{\trho\trho}}{\sqrt{1+g^{\trho\trho}g_{\theta\theta}(\partial_\trho
\theta)^2}}\,.\label{tense}\eeq
Now if we wish to calculate the effective tension for a bundle of
strings smeared out of the $x^i$-directions with density $\nq$, the
same calculation would apply since eq.~\reef{actionp3} is precisely
the fundamental string action. However, these strings would lie
vertically along the axis and so we would evaluate eq.~\reef{tense}
with $\partial_\trho \theta=0$. Hence for a cusp with any non-zero
$\partial_\trho \theta$, the effective tension \reef{tense} is less
than that of the vertical strings. Hence none of these
Minkowski-like embeddings can achieve an equilibrium with the
attached strings for any finite value of $\partial_\trho
\theta$.\footnote{The same conclusion applies for the general
Dq/Dp-brane configurations discussed in subsection \ref{new2}.} We
might consider these configurations as the initial data in a
dynamical context. Then, given the results above, we see that the
strings will pull the brane down the axis to the horizon -- a
similar discussion appears in a different context in \cite{cusp}. In
any event, we will not consider any of these Minkowski-like
embeddings in the remainder of our analysis.

Now let us consider the black hole embeddings that arise from
eq.~\reef{actionp3}. In fact, the equations resulting from this
action were studied as (a special case of) the string configurations
describing Wilson loops in the AdS/CFT \cite{wilson}. In general
these solutions are loops which begin and end at large $\trho$.
Hence these are inappropriate in the present context.\footnote{If we
use only a portion of these solutions, \ie the configuration is
cut-off before reaching the loop's minimum $\trho$, the profile
describes the cuspy configurations discussed above.} In this
context, at finite temperature, there is another class of string
configurations, namely strings that fall straight into the horizon,
which display the screening of the quark-antiquark potential. Using
this experience, we conclude that the {\it only} solutions for
eq.~\reef{actionp3} which reach the horizon will be the constant
configurations $\theta=\theta_0$. Hence, as we saw in the near
horizon analysis, the black hole embeddings near the $\theta=0$ axis
are long narrow cylinders of constant (angular) cross-section.

One should ask how far out these constant profiles are valid as
approximate solutions of the full equations derived from
eq.~\reef{actionp2}. The approximation that allowed us to derive
eq.~\reef{actionp3} required $\tilde{d}^{1/3}\gg\rho\sin\theta$,
assuming $\rho\gg1$. Hence the constant solutions $\theta=\theta_0$
should remain approximate solutions out to $\rho_{\rm
transition}\sim\tilde{d}^{1/3}/\theta_0$ for small $\theta_0\ll1$.
Beyond this radius we expect the profile should expand out and
approach an asymptotically flat brane. However, we can push this
transition out to an arbitrarily large radius by taking
$\theta_0\rightarrow0$. This again suggests that with $d\ne0$, there
are D7-brane embeddings which reach the horizon no matter how far
the (asymptotic) brane is from the black hole. We will verify this
result with numerical investigations of the full solutions for the
action \reef{actionp2} in the next subsection.

Our analysis of the static D7-brane profiles near $\chi\sim1$ have
confirmed the idea that the embeddings develop a narrow spike that
behaves like a bundle of strings stretching between the asymptotic
brane and the black hole. It is interesting to extend this idea
further by investigating the dynamical properties of these spikes.
As a step in this direction, let us consider our framework with the
more general ansatz: $\chi(\trho,t)$ and $A_t(\trho)$.\footnote{The
symmetries of the problem ensure that this ansatz leads to a
consistent solution.}  After a straightforward calculation the
Legendre-transformed action becomes
\beq \tids=-T_\mt{D7}\int d^8 \sigma\frac{f}{(2\ft)^{1/2}} 
\sqrt{1+\trho^2 (\partial_\trho \theta)^2-\frac{2L^4}{\trho^2}
\frac{\ft}{f^2}(\pa_t\theta)^2}
\left[\frac{d^2}{(2\pi\ls^2T_\mt{D7})^2}+
\frac{\nf^2}{8}\trho^6\tilde{f}^3\sin^6\theta\right]^{1/2}
\labell{actionp22} \,. \eeq
As above, we restrict our attention to the embeddings when they are
very close to the axis $\theta \simeq0$. In this regime, the second
contribution in the last factor can be neglected and
eq.~\reef{actionp22} becomes
\beq \tids \simeq - \nq V_x \frac{1}{2\pi \ls^2} \int dt\,d\trho\,
\frac{f}{(2\ft)^{1/2}}\sqrt{1+\trho^2 (\partial_\trho
\theta)^2-\frac{2L^4}{\trho^2} \frac{\ft}{f^2}(\pa_t\theta)^2} \,.
\labell{actionp33} \eeq
Once again we recognize this result as the Nambu-Goto action for a
bundle of fundamental strings stretching in the $\trho$-direction
with dynamical fluctuations in the $\theta$-direction. Hence we
are beginning to see that not just the static properties of the
spikes, such as the tension, but also their dynamical spectrum of
perturbations matches that of a collection of strings; similar
results have been seen for the dynamics of the BIon spikes on branes
in asymptotically flat spacetime \cite{peet}. In this sense we see that,
although no fundamental strings are initially manifest, the D7-brane
spectrum still captures the presence of these strings. This is a
satisfying result since these strings stretching between the horizon
and the asymptotic D7-branes are dual to the quarks in the field
theory, for which we are turning on chemical potential $\mu$. It
would be interesting to investigate these issues in more detail.

\subsection{Numerical embeddings}

We now return to the detailed analysis of the D7-brane embeddings in
the black D3-brane background. In general, it is not feasible to solve
analytically eq.~\reef{chiEom}, which
determines the profile $\chi(\rho)$, so we resorted to numerics.  We numerically
integrated eq.~\reef{chiEom}, specifying boundary conditions on the
horizon $\rho_\mt{min}=1$: $\chi(1)=\chi_0$ for various $0\leq
\chi_0 <1$ and $\partial_\rho \chi|_{\rho=1}=0$. In order to compute
the constants $m,c$ corresponding to each choice of boundary
condition at the horizon, we fitted the solutions to the asymptotic
form \reef{asymptD7}. Several representative D7-brane profiles are
depicted in figure \ref{embeddings}. In particular, we see
explicitly here the formation of long narrow spikes reaching down to
the horizon as $\chi_0$ approaches 1 (or $R$ approaches 1 on the
horizon).
\FIGURE{
 \includegraphics[width=0.45 \textwidth]{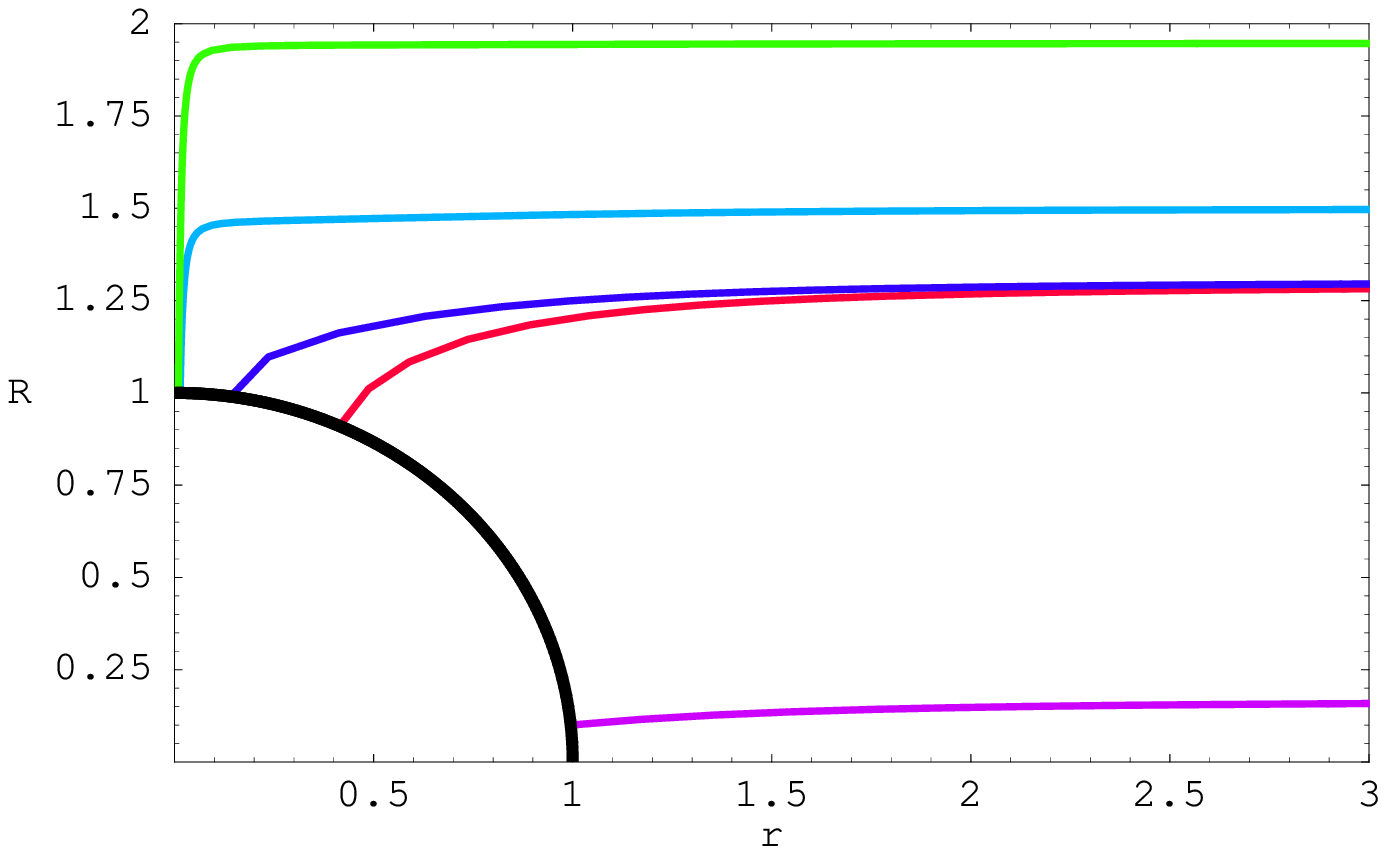}
\caption{Profiles of various D7-brane embeddings in the D3-brane
background for $\tilde{d}=10^{-4}/4$. The black circle represents
the horizon.} \label{embeddings}}

We can make the appearance of these spikes quantitative here by
examining how varying boundary the condition $\chi_0$ changes the
quark mass $m$ -- recall that the latter is proportional to the
distance which the branes reach along the vertical axis of figure
\ref{embeddings}. Figure \ref{mVsR_0} shows plots of $m$ versus
$\chi_0$ for $\tilde{d}= 10^{-4}/4$ and $1/4$. Note that in both
cases, as $\chi_0 \to 1$, the quark mass is diverging. Hence with
$\tilde{d}\ne0$, there are D7-brane embeddings which reach the
horizon no matter how large the (asymptotic) separation between the
brane and the black hole becomes. Since $m\propto \mq/T$ as shown in
eq.~\reef{mc}, $m\rightarrow\infty$ corresponds to $T\rightarrow0$
for a fixed quark mass $\mq$. Hence the previous result is
equivalent to saying that the D7-branes intersect the horizon for
all values of $T$ when $d\ne 0$. Contrast this with the
$\tilde{d}=0$ case, where embeddings of the D7-branes which
intersect the horizon (\ie black hole embeddings) only existed above
some minimum temperature \cite{prl,long}. At low temperatures the
D7-branes were described by embeddings which smoothly closed off
above the horizon (\ie Minkowski embeddings). For nonzero chemical
potential or nonzero baryon density, there are black hole embeddings
corresponding to all temperatures in the gauge theory. For small
temperatures, or large quark mass, most of the brane is very far
away from the horizon with only a very thin long spike extending
down to touch the horizon. Far from the black hole, this embedding
would look very much like a Minkowski embedding in the low
temperature phase of $\tilde{d}=0$. It differs only by the narrow
spike going down to touch the horizon.
\FIGURE{
 \includegraphics[width= \textwidth]{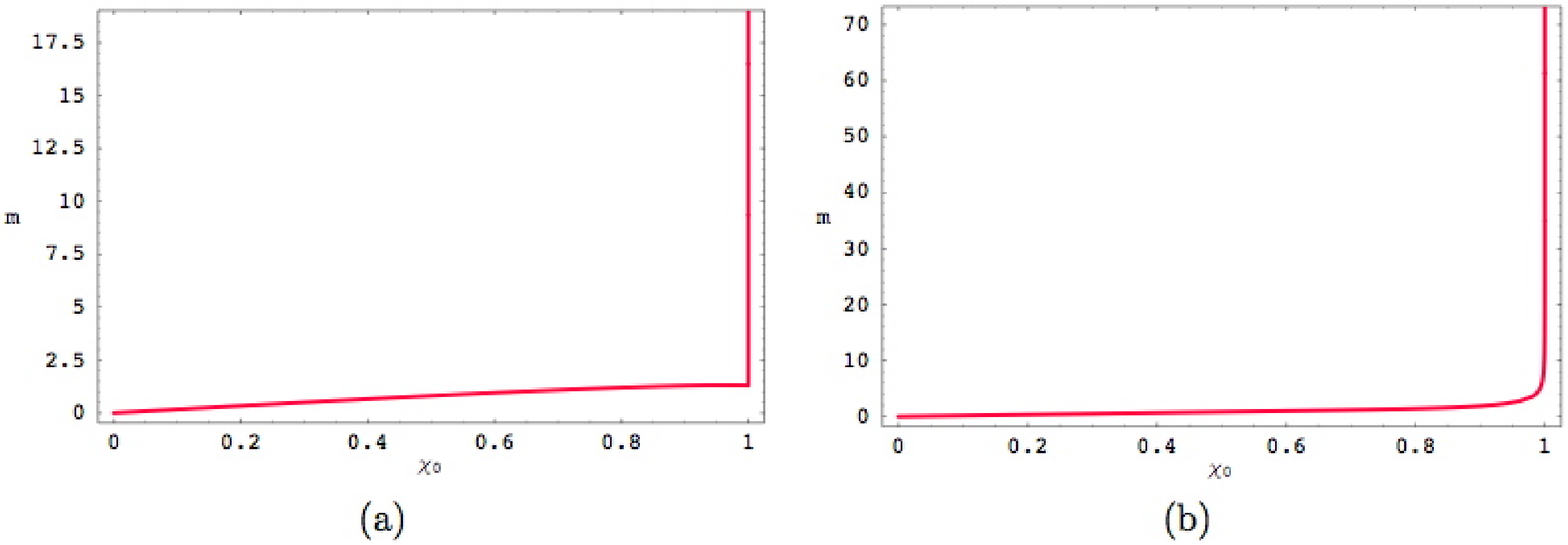}
\caption{Quark mass $m$ versus boundary condition $\chi_0$ on the
horizon for (a) $\tilde{d}= 10^{-4}/4$ and (b) $\tilde{d}=1/4$.}
\label{mVsR_0}}

Figures \ref{cVsTd10m6}, \ref{zoom} and \ref{cVsT} illustrate the
dependence of the quark condensate $c$ on the temperature $T$.
Several such plots of $c$ versus $T$ with varying degrees of
resolution are given in figure \ref{cVsTd10m6} for small values of
the baryon density: $\tilde{d}=0$, $10^{-6}/4$ and $10^{-4}/4$. In
the first two plots, the differences between the curves is virtually
indiscernable. In particular then, they all begin to show the
spiralling behaviour that was characteristic of the self-similar
scaling discovered for $\tilde{d}=0$ \cite{prl,long}. Of course,
section \ref{new2} argued that these spirals should persist to a
certain level at small $\tilde{d}$. Note that in the highest
resolution plot (the last one in fig.~ \ref{cVsTd10m6}), one sees that for
$\tilde{d}=10^{-4}/4$ the small scale spirals have been eliminated.
In any event, the plots in figure \ref{cVsTd10m6} explicitly
demonstrate that, for small baryon density $\tilde{d}$, the black
hole embeddings are mimicking the behaviour of both the black hole
and Minkowski branches of the theory at $\tilde{d}=0$. Hence certain
features of the physics will be continuous between the theories with
vanishing and non-vanishing baryon number density. In particular,
the spiralling or rather the multi-valuedness of $c$ indicates there
will be a first order phase transition and so the `melting'
transition found in \cite{prl} persists to small values of the
baryon density.
\FIGURE{
 \includegraphics[width= \textwidth]{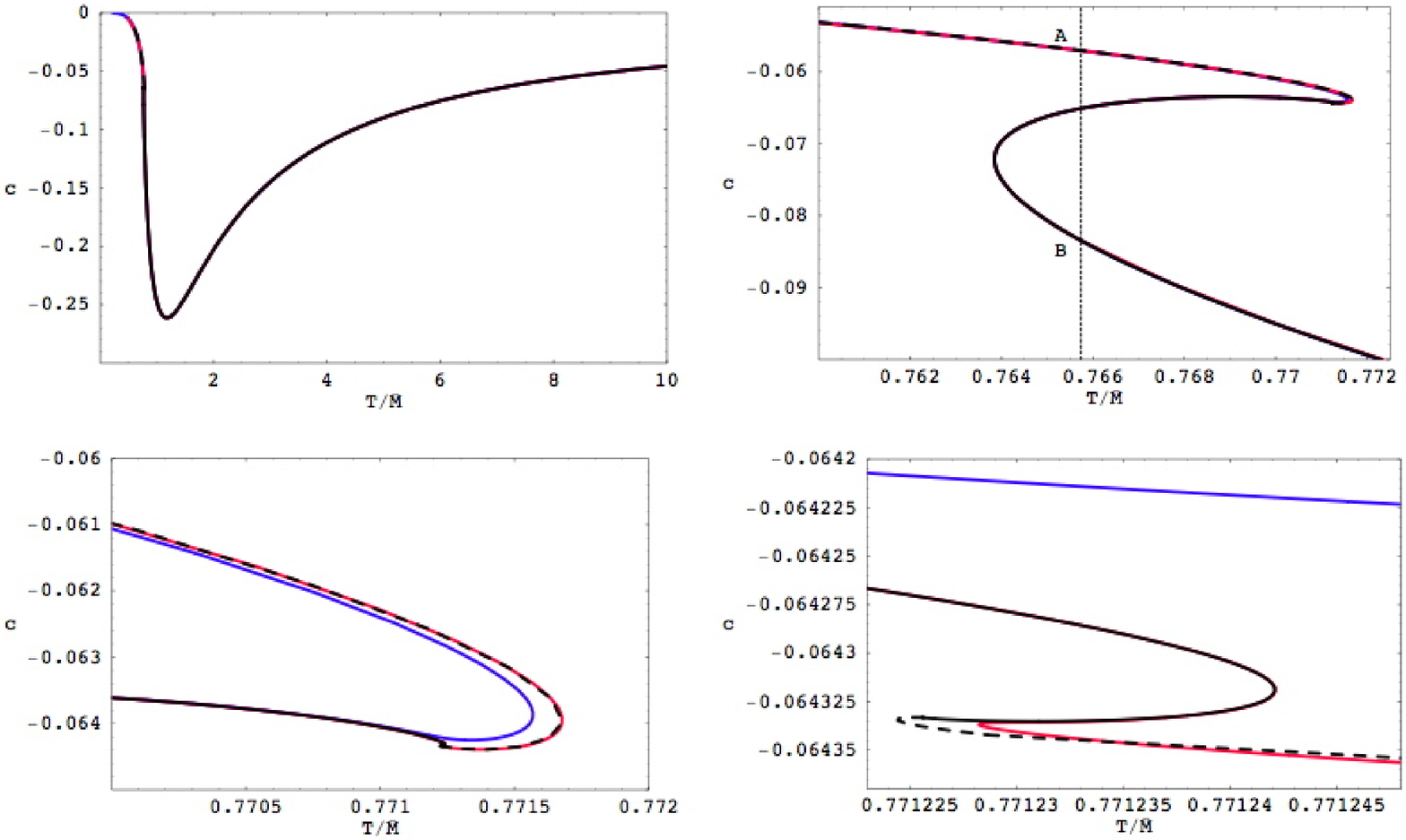}
\caption{Quark condensate $c$ versus temperature $T/\mbar$ for
$\tilde{d}=0$, $10^{-6}/4$ and $10^{-4}/4$ on the black, red and
blue curves, respectively. At low resolution, these curves are all
nearly identical and display a similar spiralling behaviour.}
\label{cVsTd10m6}}
\FIGURE{
 \includegraphics[width=0.6 \textwidth]{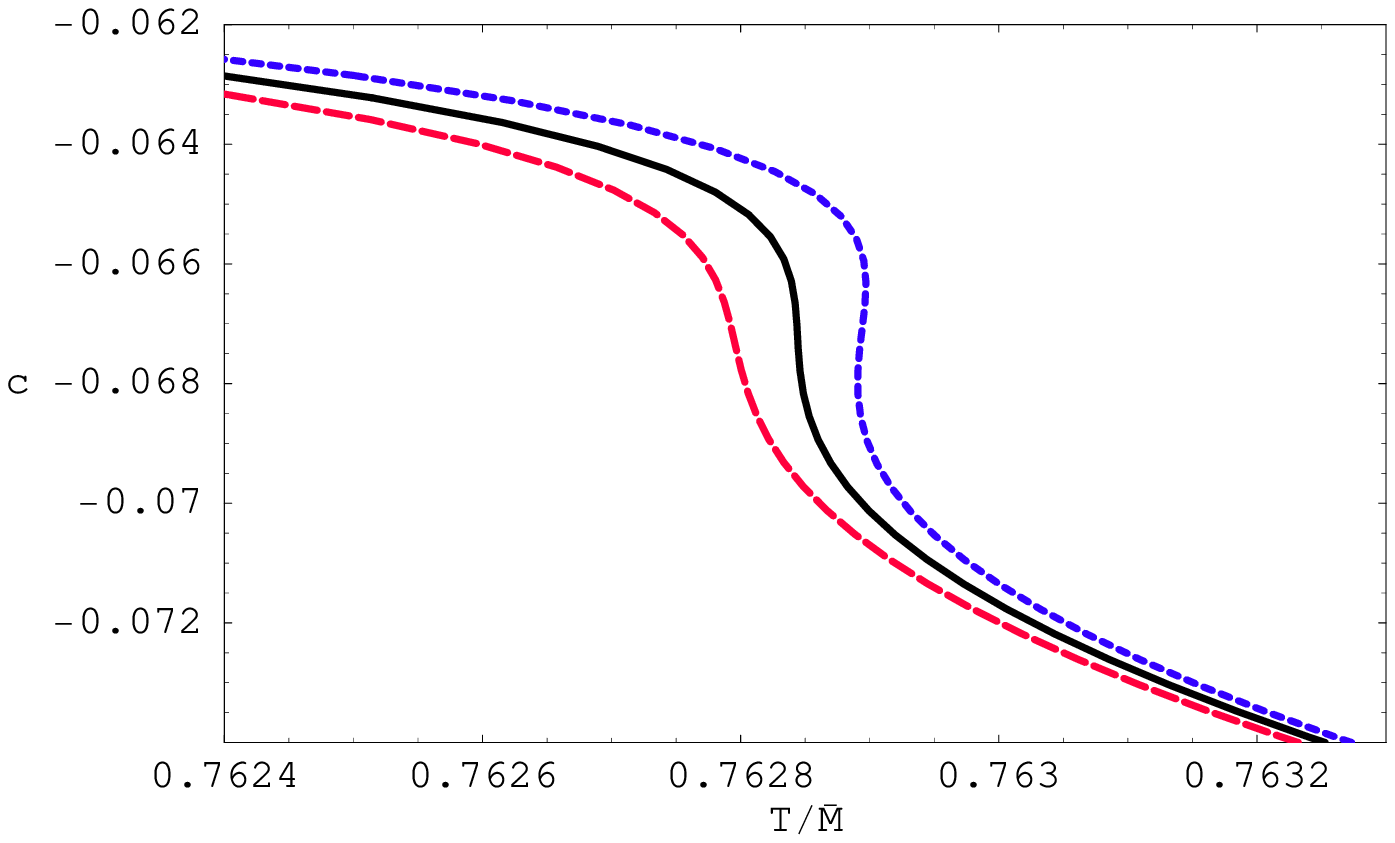}
\caption{Quark condensate $c$ versus temperature $T/\mbar$ near the
critical point. The solid black curve corresponds to the critical
baryon density $\tilde{d}^*=0.00315$. The dashed curves above (blue)
and below (red) correspond to $\tilde{d}=0.0031$ and $0.0032$,
respectively.} \label{zoom}}

As $\tilde{d}$ is increased, the self-similar, spiralling behaviour
becomes less and less pronounced and eventually $c$ becomes a
single-valued function of $T/\mbar$. To the best numerical accuracy
that we could achieve, the critical value at which the phase
transition disappears in $\tilde{d}^*=0.00315$. Figure \ref{zoom}
shows the $c$ in the vicinity of the transition around this critical
value. For $\tilde{d}=0.0031$, the curve shows a slight S-shape and
so a small first order phase transition would still occur. For the
critical value $\tilde{d}^*=0.00315$, the curve is monotonic but
with a singular slope near the center. In this case, the phase
transition would be reduced to second order. Finally for
$\tilde{d}=0.0032$, the curve is monotonic with a finite slope
everywhere and so the phase transition has disappeared.

For completeness, we also show the behaviour of the quark condensate
at much larger values of the baryon density in figure \ref{cVsT}.
Figure \ref{cVsT}a corresponds to $\tilde{d}=1/4$ where some
interesting structure still persists around $T/\mbar\sim1$, which
was where $c$ shows a minimum in figure \ref{cVsTd10m6} at smaller
densities. Figure \ref{cVsT}b corresponds to $\tilde{d}=10$, where
$c$ has become a monotonically increasing (towards zero) function of
$T$.
\FIGURE{
 \includegraphics[width= \textwidth]{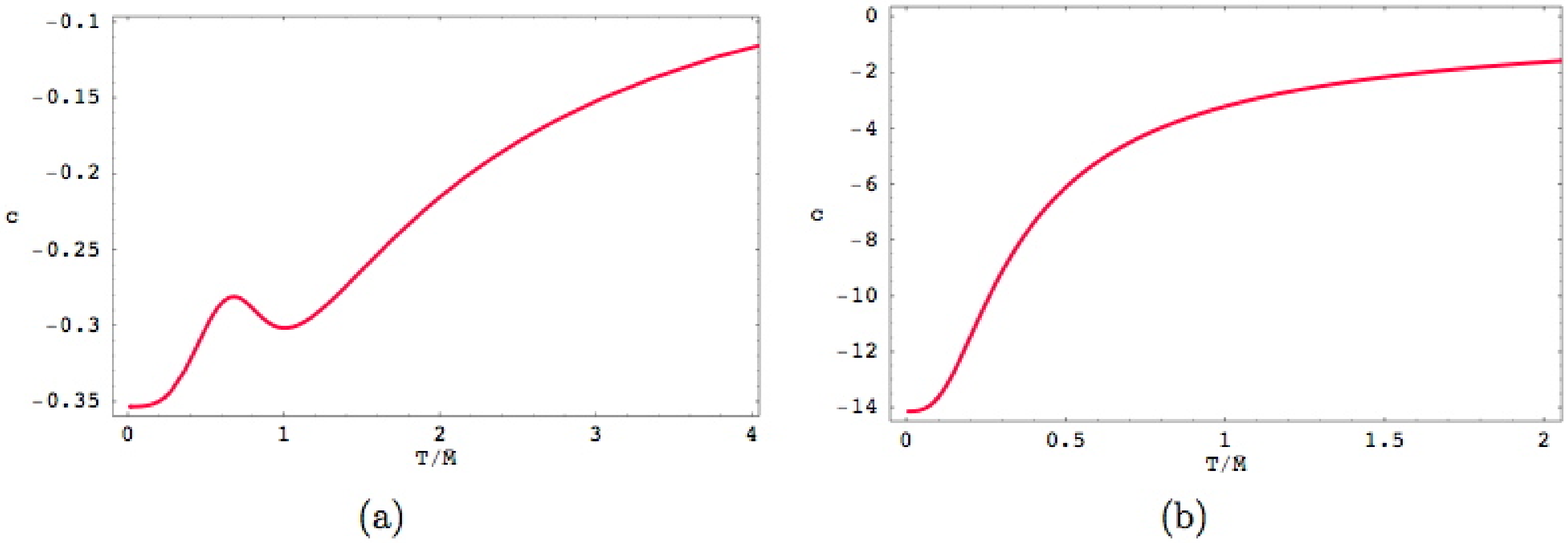}
\caption{Quark condensate $c$ versus temperature $T/\mbar$ for (a)
$\tilde{d}= 1/4$ and (b) $\tilde{d}=10$.} \label{cVsT}}

We integrated \reef{moo} numerically to solve for the chemical
potential. Plots of $\tilde{\mu}$ versus temperature all show an
apparent divergence as $T/\mbar \to 0$, as illustrated in figure
\ref{muVsT}a. However, this behaviour is misleading as we now
explain. As discussed in the previous subsections, a common feature
of the D7-brane embeddings at small temperatures is the long narrow
spike close to the $\theta=0$ axis. This spike dominates
eq.~\reef{moo} for small $T/\mbar$ and so the latter formula can be
simplified to
\beq \mu\simeq \frac{1}{\sqrt{2}2\pi\ell_s^2}\int_{u_0}^{u_0m}
d\trho\,f/\ft^{1/2}\simeq \mq\,, \labell{moo2}\eeq
where we have restored the dimensions of the chemical potential and
the radial coordinate. Hence in this limit, the chemical potential
is essentially given by the quark mass, as one might have expected.
Hence the divergence in figure \ref{muVsT}a arises simply because
$\tilde{\mu}\propto\mu/T$, as shown in eq.~\reef{for1}. This
spurious behaviour is eliminated by plotting
$\mu/\mq=\sqrt{2}\tilde{\mu}/m$, as shown in figure \ref{muVsT}b.
The latter plot exhibits the small temperature limit
$\mu/\mq\rightarrow1$ for $T$ approaching zero, as is implied by
eq.~\reef{moo2}. Note that if one calculates $\mu$ in the vicinity
of the phase transition, it shows a multi-valuedness similar to that
shown for the quark condensate above.
\FIGURE{
 \includegraphics[width=\textwidth]{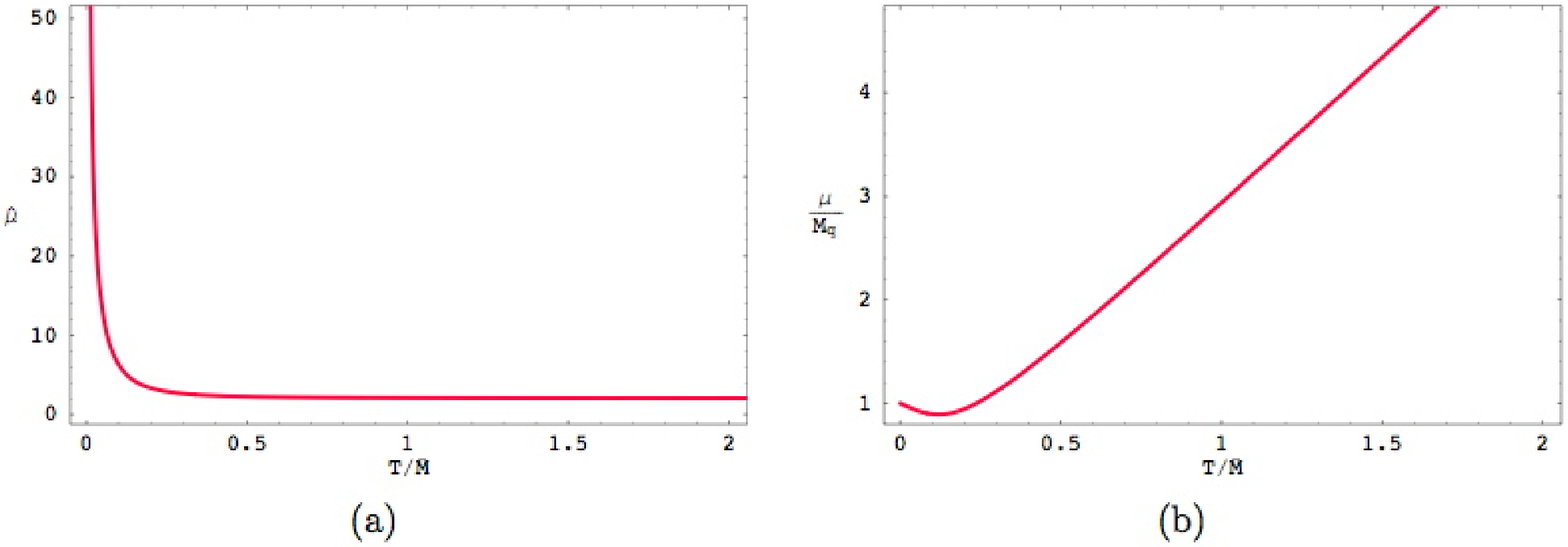}
\caption{Chemical potential for $\tilde{d}=10$ versus temperature
displayed as: (a) $\tilde{\mu}$ and (b) $\mu/\mq$.} \label{muVsT}}

\section{D7-brane thermodynamics: Free energy, entropy and stability}
\label{thermal}

We now wish to study the thermal properties of the fundamental
hypermulitplets at finite baryon number. Our holographic framework
translates this question into one of investigating the thermal
contributions of the D7-branes on the gravity side. As usual, we use
the standard technique \cite{hawk} of Wick rotating the time
direction. The Euclidean time circle of the black D3-brane
background then becomes the thermal circle in a finite temperature
path integral, and the leading contribution to the free energy is
determined by evaluating the Euclidean action. As we are interested
in the contributions of the fundamental matter, we only study the
action of the D7-branes. Although evaluating the bulk brane action
leads to a formally divergent result, the AdS/CFT correspondence
provides a prescription to remove these divergences: One introduces
a finite-radius UV cut-off and a set of boundary counterterms to
renormalise the action \cite{karch1}. This approach for the branes
is completely analogous to the same calculations which are performed
for the gravity background \cite{ct}. This holographic
renormalisation of the D7-brane action was discussed in more detail
in refs. \cite{prl,long}, which we follow closely in this section.

We begin by writing the Euclidean action for the D7-branes in terms
of dimensionless quantities, introduced in section \ref{22},
as\footnote{For simplicity, we have left $A_t$ untouched here rather
than introducing a Wick rotated potential $A_{t_\mt{E}}=-i\,A_t$. As
is  well-known, such a Euclidean potential would have to be treated
as an imaginary field in the present context because the chemical
potential and particle density must remain real constants -- see,
\eg \cite{charge}.}
%
\be I_\mt{bulk} = \int d^8 \sigma \, {\cal L}_\mt{E} = {\cal N} \int
d\rho \, \rho^3 f \tilde{f} (1-\chi^2) \sqrt{1-\chi^2+\rho^2
{\dot{\chi}}^2 - \frac{2 \tilde{f}}{f^2} (1-\chi^2)
\dot{\tilde{A_t}}^2} \,, \label{dimensionless} \ee
where ${\cal N}$ is the normalisation constant introduced in
\cite{prl,long}:\footnote{Note that this constant does not include
the three-volume $V_x$ along the gauge theory directions. Rather in
this section we will divide out these factors everywhere and so all
extensive quantities are actually densities per unit volume; for
example, \eqn{dimensionless} is the Euclidean action density.}
\beq \mathcal{N}= \frac{2 \pi^2 \nf T_\mt{D7} u_0^4}{4T} =
\frac{\lambda \nc \nf T^3}{32} \,. \eeq
The normalisation factor illustrates the fact that the leading
contributions of the fundamental matter are proportional to
$\nc\,\nf$, in accord with the large-$\nc$ counting rules of the
gauge theory. Note then that these contributions are subleading to
those of the adjoint fields which scale as $\nc^2$ -- see, for
example, the entropy density in eq.~\reef{entropy}.

As commented above, this bulk action \reef{dimensionless} contains
large-$\rho$, UV divergences. Fortunately, however, these are the
same as in the absence of the gauge field, and therefore no new
counterterms are required beyond those derived in
refs.~\cite{prl,long}, which take the form
\beq \frac{I_\mt{bound}}{\mathcal{N}}= -\frac{1}{4} \left(\rhomax^4
- 2m^2 \rhomax^2 +m^4 -4 mc \right) \,, \labell{Ibound} \eeq
where $\rhomax$ is the UV cut-off. The regularised D7-brane action
is then $\ide=I_\mt{bulk}+I_\mt{bound}$. It can most simply be
written as:
\beq \frac{\ide}{\N} = G(m) - \frac{1}{4}
\left[(\rhomin^2-m^2)^2-4mc \right] \,,
\labell{acta}\eeq
where $G(m)$ is the integral:
%
\beq G(m) = \int_{\rhomin}^{\rhomax} d\rho \left( \rho^3 f \tilde{f}
(1-\chi^2) \sqrt{1-\chi^2+\rho^2 {\dot{\chi}}^2 - \frac{2
\tilde{f}}{f^2} (1-\chi^2) \dot{\tilde{A_t}}^2} - \rho^3 + m^2 \rho
\right) \,. \eeq
The limit  $\rhomax \to \infty$ may now be taken, since this
integral converges.

As usual, we wish to identify the action with a thermodynamic free
energy. However, in the present case, there are various
possibilities depending on the ensemble under consideration, \ie
the Gibbs free energy for the grand canonical ensemble with fixed
$\mu$ and the Helmholtz free energy for the canonical ensemble with
fixed $\nb$. Now experience with similar calculations for charged
black holes, \eg \cite{charge}, suggests that the Gibbs free energy
is given by the Euclidean action while the Helmholtz free energy is
associated with the Legendre transform of $\ide$. Since we wish to
work with fixed charge, we would want to work with the latter.

In the following, we confirm the above expectations. Using the
equations of motion, the variation of the action reduces to a
boundary term:
%
\beq \delta \ide = \left[\frac{\partial \mathcal{L}_\mt{E}}{\partial
\dot{\chi}} \delta \chi + \frac{\partial
\mathcal{L}_\mt{E}}{\partial \dot{\tilde{A}}_t} \delta \tilde{A}_t
\right]^{\rhomax}_{\rhomin}. \eeq
Combining this with the variation of the boundary action
$I_\mt{bound}$ \reef{Ibound} yields
\beq
\delta \ide = -2 \N c \, \delta m -\frac{\nq}{T} \delta \mu
\eeq
where $\nq$ was defined in \reef{ad}. Recalling that $m=\mbar/T$ we
see that the natural thermodynamic variables of the Euclidean action
are the temperature $T$ and the chemical potential $\mu$. Hence we
must identify $\ide = \beta W$, where $W(T, \mu)$ is the
thermodynamic potential in the grand canonical ensemble, namely the
Gibbs free energy.

Since we wish to work at fixed charge density, \ie in the canonical
ensemble, we perform a Legendre transformation by defining
\beq \tide = \ide
+\frac{\nq \, \mu}{T} \,, \ee
which of course is a function of the
temperature and the charge density:
\beq \delta \tide = -2 \N \, c
\, \delta m + \frac{\mu}{T} \, \delta \nq \,. \eeq
We thus identify $\tide = \beta F$ where $F(T,\nq)$ is the Helmholtz
free energy.

The  bulk part of $\tide$ is of course  the Euclidean analogue of
\eqn{actionp}:
\be
\frac{\tilde{I}_\mt{bulk}}{\mathcal{N}}= \int d\rho\, \rho^3 \, f
\tilde{f} (1-\chi^2) \sqrt{1-\chi^2 +\rho^2 \dot{\chi}^2}
\left[1 +\frac{8 \tilde{d}^2}{\rho^6 \tilde{f}^3
(1-\chi^2)^3} \right]^{1/2} \,.
\labell{EuclidAction}
\ee
Since the divergences of this bulk action are the same as those of the
$\tilde{d}=0$ case, the analogous expression to eq.~\reef{acta} is now
\beq \frac{\tide}{\N} = \widetilde{G}(m) - \frac{1}{4}
\left[(m^2-1)^2-4mc \right] \,, \labell{actb}\eeq
where $\widetilde{G}(m)$ is the integral:
\beq \widetilde{G}(m) = \int_{1}^{\infty} d\rho \left[\rho^3 f
\tilde{f} (1-\chi^2) \sqrt{1-\chi^2+\rho^2 \dot{\chi}^2}
\left(1+\frac{8 \dd^2}{\rho^6 \tilde{f}^3 (1-\chi^2)^3}
\right)^{1/2}-\rho^3+m^2 \rho \right]\,. \labell{wg}\eeq
In both of these expressions, we have replaced $\rho_{min}=1$ since
all of the embeddings which we consider terminate at the horizon.

We evaluated the free energy numerically for various $\dd$ and
representative results are given in figures \ref{LegVsT} and
\ref{LegVsTd01}. The behaviour of the action versus temperature in
figure \ref{LegVsT} for $\dd= 10^{-4}/4$ is nearly identical to that
for $\dd=0$ -- we refer the interested reader to compare with the
plots presented in \cite{prl,long}. The results for $\dd= 10^{-4}/4$
are typical for small values of $\dd$ with the classic `swallow
tail' shape. Of course, the crossing point of the two branches
coming in from small and large $T$ marks the temperature of the
phase transition. By varying $\dd$, one can then map out the phase
diagram shown above in figure \ref{phase-diagram}. A more detailed
diagram is shown here in figure \ref{phazze}. We see here that the
first order phase transition occurs along a segment starting at
$T_\mt{fun}/\mbar=.7658$  at $\dd=0$ and ending at the critical
point at $T^*_\mt{fun}/\mbar=.7629$ and $\dd^*=0.00315$.
\FIGURE{
 \includegraphics[width= \textwidth]{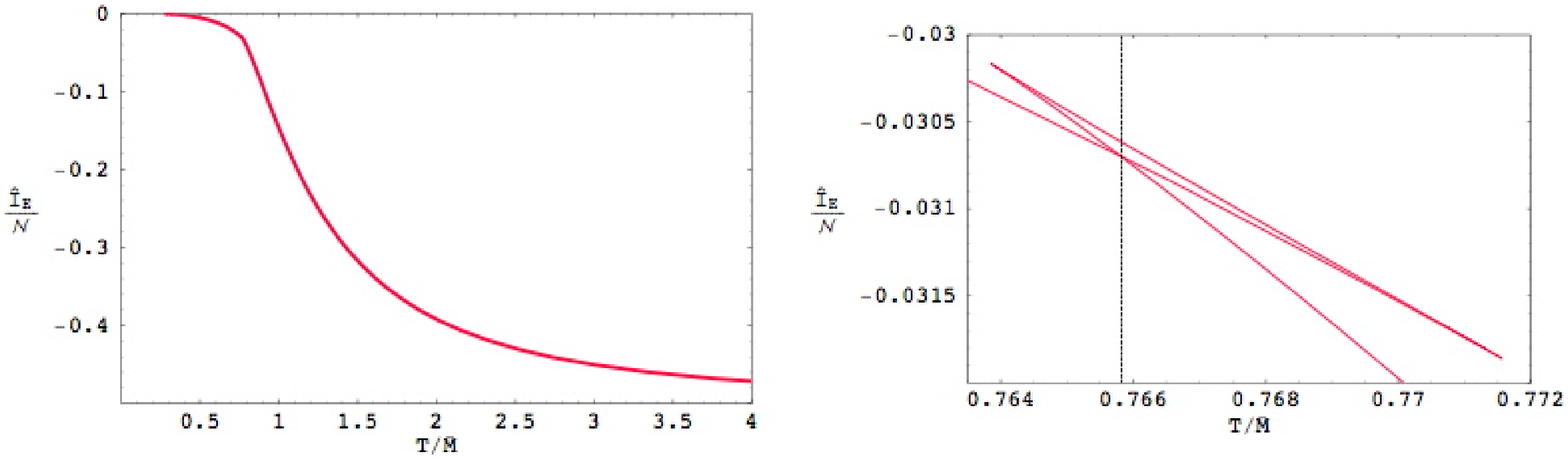}
\caption{Legendre transform of the action, $\tilde{I}_\mt{D7}$, versus
temperature for $\tilde{d}= 10^{-4}/4$. The phase transition
temperature is denoted by the dotted vertical line in the second plot.}
\label{LegVsT}}
\FIGURE{
 \includegraphics[width=0.6 \textwidth]{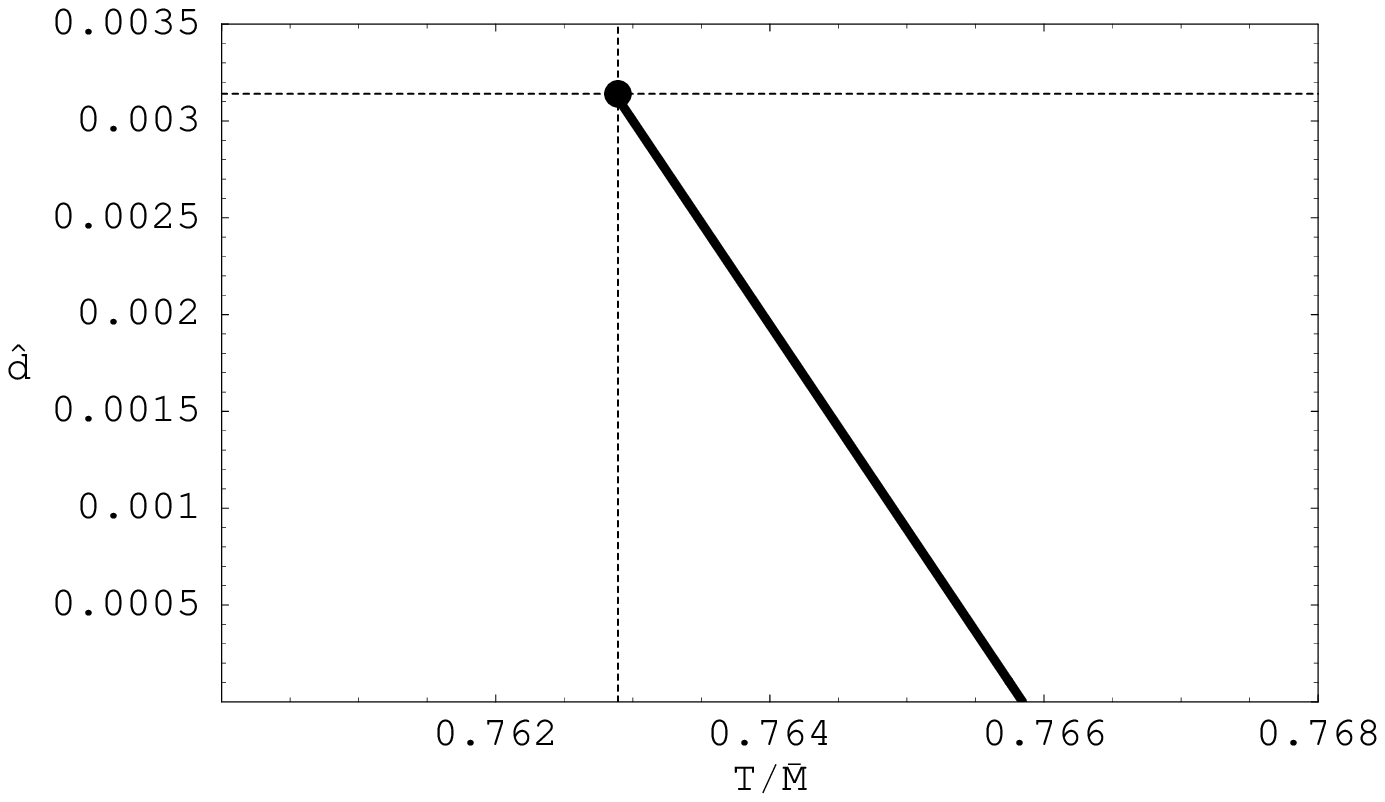}
\caption{Phase diagram: Baryon density $\dd$ versus temperature
$T/\mbar$.} \label{phazze}}

For completeness, we show some representative plots for large values
of $\dd$ in figure \ref{LegVsTd01}, where there is no crossing and
no phase transition. Note that these plots show an apparent
divergence as $T\ra0$ but this is a spurious effect in analogy to
the discussion of the plots for the chemical potential. This
artifact is actually present in all of the free energy plots but the
width becomes very narrow at small $\dd$.
\FIGURE{
 \includegraphics[width= \textwidth]{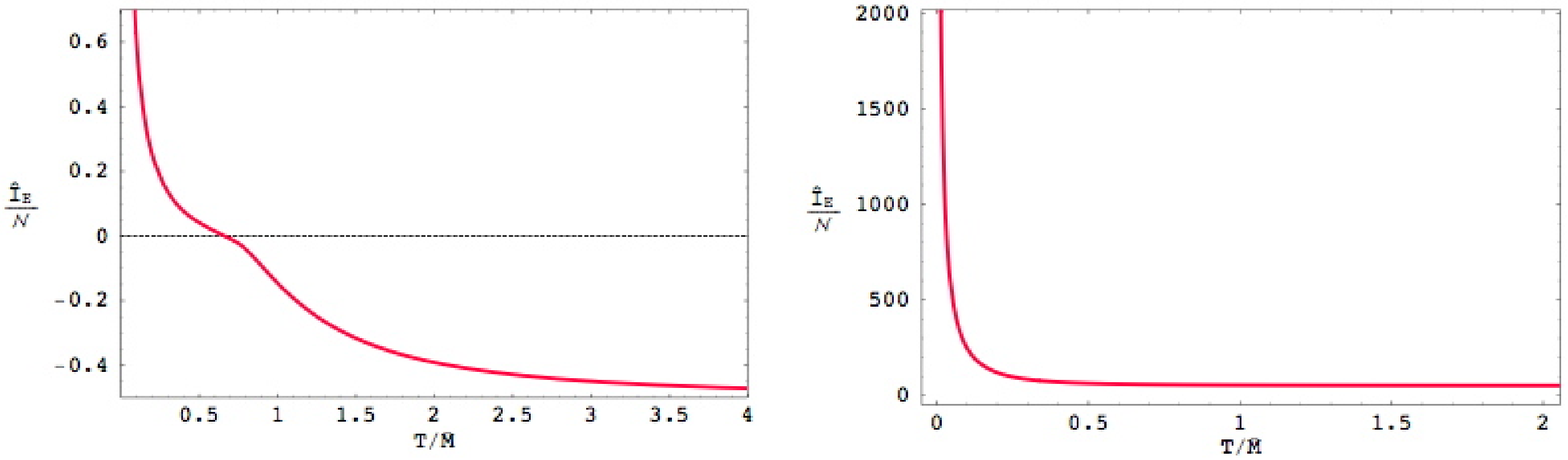}
\caption{Legendre transform of the action, $\tilde{I}_\mt{D7}$, versus
temperature for (a) $\tilde{d}= 10^{-1}/4$ and (b)  $\tilde{d}= 10$.
There is no phase transition for these values of $\tilde{d}$.}
\label{LegVsTd01}}

We now turn to the entropy density. This can be obtained by differentiating the
Helmholtz free energy density $F(T, d)= T\tide$ with respect to $T$ as
\beq S = - \frac{\partial F}{\partial T} = -\pi L^2 \frac{\partial F}{\partial u_0} \,,
\labell{entropy3} \eeq
where we used the relation $u_0 = \pi L^2 T$. Following the
calculations described in \cite{long}, one must carefully consider
all of the implicit $u_0$ dependence in \reef{actb}. The only new
contribution comes here from the appearance of $\tilde{d}$ in
\reef{wg} since from (\ref{redef}), we can see that
\begin{equation}
\frac{\partial \tilde{d}}{\partial {u_0}} = -\frac{3}{u_0} \tilde{d}\,.
\end{equation}
Gathering all the contributions, the entropy can be expressed as
\beq {S \over \N} = -4 \widetilde{G}(m)+24\tilde{d}^2 H(m) +(m^2
-1)^2 -6mc. \labell{sss6} \eeq
Here we have defined the integral
\begin{equation}
 H(m) = \int_1^\infty d\rho \
  \frac{f\sqrt{1-\chi^2 +\rho^2 \dot{\chi}^2}}{\rho^3\tilde{f}^2(1-\chi^2)^2}
  \left[1 +\frac{8 \tilde{d}^2}{\rho^6 \tilde{f}^3 (1-\chi^2)^3}
\right]^{-1/2}\,.
\end{equation}
Comparing this expression to eq.~\reef{moo}, we see that
$H=\tilde{\mu}/2\tilde{d}$. Hence we may write the final result as
\beq {S \over \N} = -4 \widetilde{G}(m)+12\,\tilde{d}\tilde{\mu}
+(m^2 -1)^2 -6mc\,. \labell{sss7}\eeq
We evaluated the entropy  numerically for various $\dd$ and some
typical results are given in figs. \ref{entropyd10m4} and
\ref{entropyBigd}. The behaviour of the action versus temperature in
figure \ref{LegVsT} for $\dd= 10^{-4}/4$ is nearly identical to that
for $\dd=0$. In particular, near the phase transition point, the
curve is multi-valued because there are several embeddings with the
same values of $\dd$ and $T/\mbar$. We refer the interested reader
to compare with the plots presented in \cite{prl,long}. Figure
\ref{entropyBigd} shows the behaviour of the entropy for larger
values of $\dd$ beyond the critical point.
\FIGURE{
 \includegraphics[width= \textwidth]{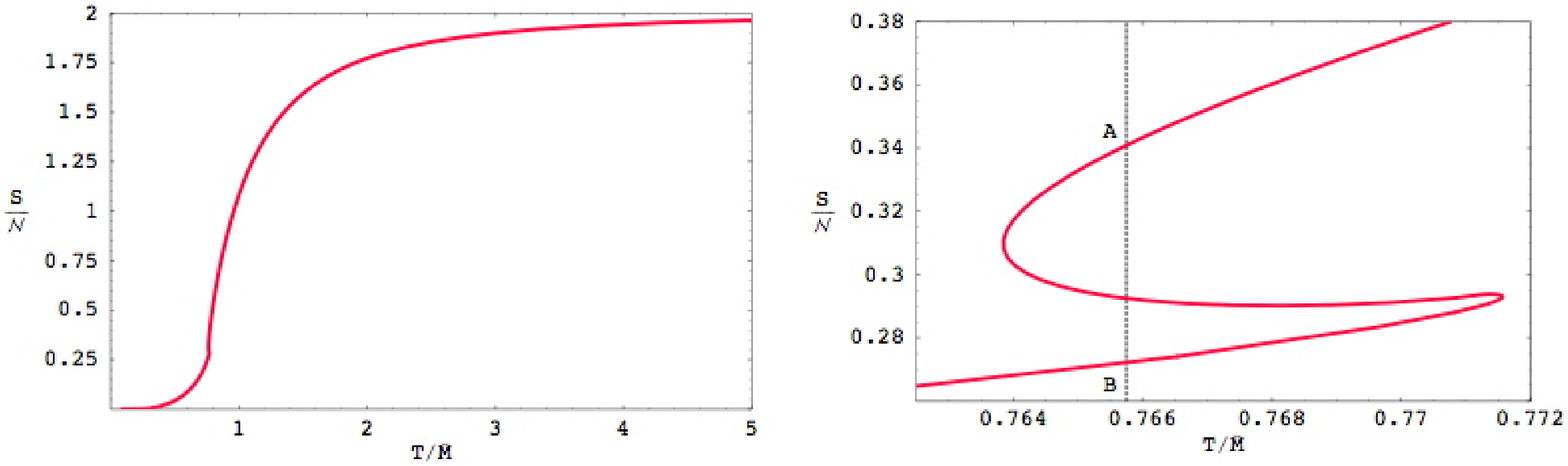}
\caption{The entropy $S/\N$ versus temperature $T/\mbar$ for
$\tilde{d}= 10^{-4}/4$.  The position of the phase transition is
marked by the dotted vertical line in the second figure. }
\label{entropyd10m4}}
\FIGURE{
 \includegraphics[width= \textwidth]{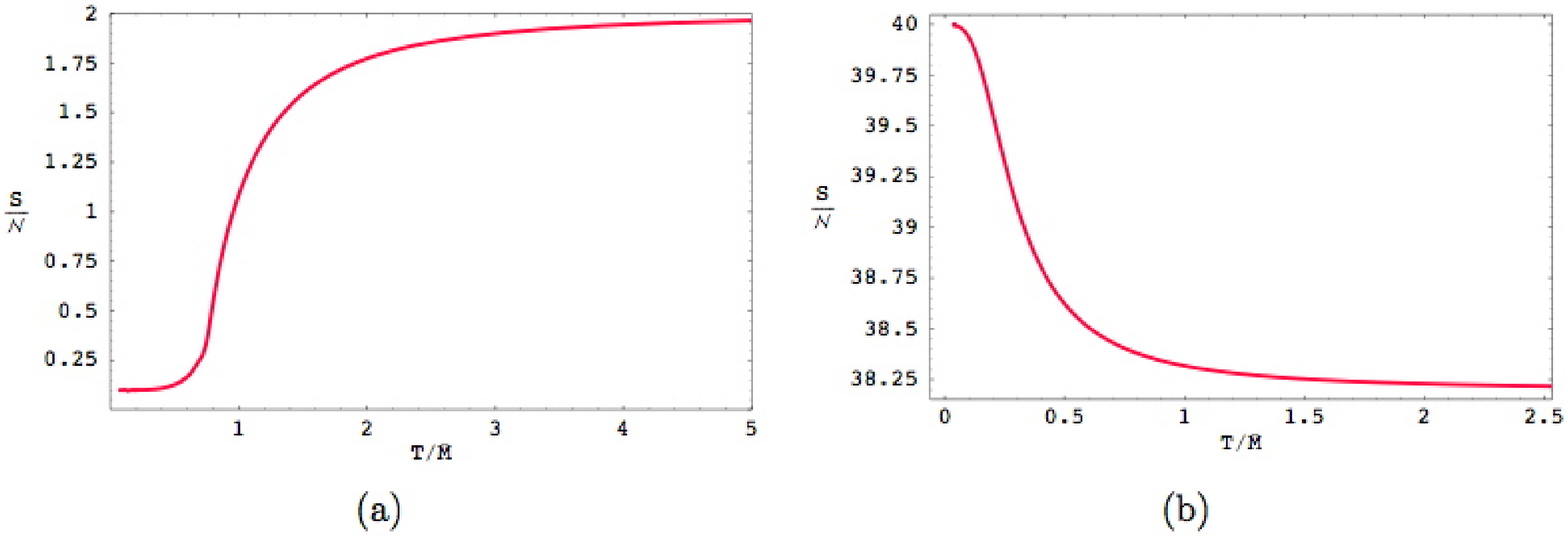}
\caption{The entropy $S/\N$ versus  temperature $T/\mbar$ for (A)
$\tilde{d}= 10^{-1}/4$ and (b) $\tilde{d}=10$.} \label{entropyBigd}}

The thermodynamic identity $E=F+T\,S=T(\tide +S)$ allows us to
determine the contribution of the D7-brane to the energy density:
\beq \frac{E}{\mathcal{N}T} = -3\widetilde{G}(m)+12\,\dd \tilde{\mu}
+\frac{3}{4}\left[(m^2 -1)^2 -\frac{20}{3}mc \right]\,.
\labell{enerve}\eeq
While we did calculate $E$ for many values of $\dd$, we do not
present any plots here as qualitatively their behaviour is similar
to that in the plots of the entropy.

Finally, we turn to the thermodynamic stability of the system. There
are various ways to write the requirements for the intrinsic
stability of our fixed-charge ensemble. We investigated stability
here with the conditions:
\beq \frac{\pa S}{\pa T}>0 \ ,\qquad \frac{\pa \mu}{\pa n}>0\ .
\labell{test}\eeq
The first one requires that the system be stable against
fluctuations in energy and seems to be satisfied everywhere. The
second constraint for electrical stability is more interesting, as
we found that it was not satisfied for all $\dd$ and $T$. Our
investigations of the region of instability remain preliminary, but
figure \ref{phase-diagram} roughly illustrates the extent of the
unstable zone as the shaded (red) region. In particular, the line of
the phase transition seems to be part of the boundary of the
unstable region between $T^*_\mt{fun}$ and $T_\mt{fun}$. This would
indicate that the black hole embeddings do not correctly describe
the true ground state in this small region and in particular, just
below the phase transition. We hope to return to this matter in the
future.
We comment more on the implications of the instability in
the discussion section below.

\section{Discussion}
\label{discuss}

Ref.~\cite{prl} identified a universal, first order thermal phase
transition in holographic Dp/Dq systems. This was characterised by a
jump of the Dq-branes between a Minkowski embedding and a black hole
embedding in the background of the black Dp-branes. In the gauge
theory this transition is associated to the melting of the mesons.

Here we have shown that Minkowski embeddings become inconsistent at
any finite baryon (or equivalently, quark) number density. The
physical reason is that a non-zero density which is dual to a
worldvolume electric field translates into a finite number of
strings being dissolved into the Dq-branes. Hence the brane is not
allowed to close off smoothly as the strings cannot simply
terminate. We considered the possibility of Minkowski-like
embeddings where the branes close off above the horizon and external
fundamental strings are attached at this point and extend down to
the horizon. However, examining the forces between the cusp in the
brane embedding and the external strings, one finds that no
equilibrium configuration is possible. Rather the strings would pull
the tip of the brane down to meet the horizon. We note here though
that this is not the only possibility for a Minkowski-like
embedding. One must simply attach a source for the strings and one
obvious alternative for such source is the baryon vertex
\cite{baryon-vertex}. In a Dp-brane background, the baryon vertex
consists of a D(8--p)-brane wrapping the internal $S^{8-p}$. Hence
it may be that there is a family of Minkowski-like embeddings, where
a gas of baryons absorbs the strings dissolved on the probe branes.
It would be interesting to investigate this possibility further.

On the other hand, we did find that with any non-zero baryon density
$\nb$, black hole embeddings where the Dq-branes intersect the
horizon exist for all values of the temperature. In contrast, such
embeddings do not exist below a certain temperature for $\nb=0$ and
the system must be described by a Minkowski embedding beyond this
point. In any event, we focused here on studying the behaviour of
the black hole embeddings at finite $\nb$ in the specific example of
the D3/D7 system. Our results indicate that the physics is
essentially continuous around $\nb=0$. The reason is that black hole
embeddings with very small $\nb$ mimic the behaviour of both $\nb=0$
Minkowski embeddings and $\nb=0$ black hole embeddings. Moreover,
the near-horizon analysis strongly suggested that the universal
phase transition found in \cite{prl} should persist for sufficiently
small baryon densities, but that it should cease to exist above some
critical value $\nb=\nb^*$. This was confirmed by our detailed
numerical analysis for the D3/D7 system. We emphasize though that
the transition at small baryon density occurs between two black hole
embeddings.

At zero baryon number density, the spectrum on Minkowski embeddings
consists of a gapped, discrete set of stable mesons (in the
large-$\nc$, strong coupling limit), together with stable, massive,
free constituent quarks \cite{prl,long}. Instead, mesons on black
hole embeddings have melted and and the spectrum is continuous
and gapless. In fact, little evidence of the previously stable
states remains in this continuous spectrum \cite{sound2}. In
addition, constituent quarks are massless. In the presence of a
non-zero baryon density, all embeddings are of black hole type and
hence no strictly stable mesons exist. Note, however, that the decay
width is very small if the quark mass is very large, or if the meson
is very heavy. Indeed, the decay width of a meson is proportional to
the support of its wave function on the near-axis region where the
spike attaches to the branes. This region becomes small as the quark
mass increases. Alternatively, the peak of the meson wave function
occurs further and further away from the axis as the meson mass
increases -- which, for fixed quark mass, can be achieved by, for
example, increasing the meson radial quantum number. We plan to
study these issues in more detail elsewhere. 

Similarly, it may seem that the free constituent quarks represent a
puzzle in this framework. Recall that the dual gauge theory is
deconfined and so free quarks should play a role, in particular
since we introduce a chemical potential. The analysis at $\nb=0$
suggests that at least at low temperatures a constituent quark is
dual to a string extending from the horizon to the brane (at large
radius). However, at finite $\nb$, our embeddings are all of the
black hole type and so if we attach such a string to the brane, it
will quickly slip away behind the horizon. Hence the puzzle is: How
do the D7-branes capture the physics of a gas of constituent quarks
at low temperatures when there are no stable excitations
corresponding to macroscopic strings?

Of course, the resolution of this puzzle is provided by the analysis
in subsection \ref{new}. The near-horizon analysis of the Dp/Dq
system suggested that, for any value of the baryon density, there
should exist Dq-brane embeddings which resemble closely Minkowski
embeddings everywhere except for a long thin spike stretching all
the way down to the black Dp-branes horizon. This was confirmed for
the D3/D7 case by our numerical results, which demonstrate that such
embeddings correspond to large quark masses (or low temperatures).
Further, we showed that not only do these spikes match the tension
of a bundle of fundamental strings, but also their dynamics. Hence
these spikes provide a brane realisation of the desired gas of
constituent quarks. Since the fields describing the D7-branes are
dual to meson operators (\ie operators with $\nq=0$) in the gauge
theory, we may say that, in a very precise sense, quarks are being
built out of mesons here, in the limit of large quark masses.

In considering the discussion above, one must remember that part of
our phase diagram \ref{phase-diagram} corresponds to unstable
embeddings. In particular, the line of the phase transition seems to
be part of the boundary of the unstable region. This would indicate
that the black hole embeddings do not correctly describe the true
ground state of the phase immediately below the phase transition.
Hence while one should not doubt the existence of a phase
transition, the precise location of the transition can be called
into question. Recall however, that for small $\dd\ne0$ the
behaviour of the black hole embeddings matched everywhere the known
behaviour of the system with $\dd=0$ very closely, as illustrated in
fig.~\ref{cVsTd10m6}. Hence we expect that the true line of phase
transitions must be very close to that indicated in
fig.~\ref{phazze} for small $\dd$ but it may deviate to the right at
larger values of $\dd$. We also reiterate that we are still refining
our results on the boundary of the unstable region 
and that fig.~\ref{phase-diagram} only gives a qualitative
representation beyond $T^*_\mt{fun}$. It may also be that the region
below the phase transition line very close to $T_\mt{fun}$ is
stable.

The instability arises in the region where
\be \left( \frac{\pa \mu}{\pa
\nb} \right)_T = \left(\frac{\pa^2 F}{\pa \nb^2} \right)_T <0 \,.
\label{inst} \ee
It would of course be interesting to identify what the stable
ground state is in this region. One indication comes from the nature
of the instability itself. In the region where \eqn{inst} holds the
free energy $F$ is a concave function of the baryon density, which
means that the system can lower its free energy by separating into
two phases with densities $\nb^1 < \nb < \nb^2$ such that
\be \gamma \nb^1 + (1-\gamma) \nb^2 = \nb \sac
\gamma F(\nb^1) + (1-\gamma) F(\nb^2) < F(\nb) \,. \ee
One way in which this would be realised in the gravity description
would be that it becomes thermodynamically favourable for the $\nf$
D-brane probes to distribute the $U(1)_q$ charge unequally among
constituent branes, presumably through some mechanism involving the
non-Abelian nature of their dynamics. 
This would imply that the flavour symmetry is spontaneously broken in the 
infrared. Alternatively,  such a
separation in different-$\nb$ phases may be realised by going to a spatially inhomogeneous phase where $\nb$ varies from point to
point. The Minkowski-like embeddings carrying gas of baryons may play a role in
this regime. We would also note that at this point, it is not clear
whether or not other phases or embeddings will also play a role
beyond the region of instability. In particular, we suspect that a
new phase may appear at very low temperatures.

In this paper we have concentrated on the phase structure of gauge
theories at constant temperature and charge density, namely on their
description in the canonical ensemble. It will be interesting to
consider the phase structure of these theories in the grand
canonical ensemble, \ie as a function of the temperature and the
chemical potential 
This should be particularly
interesting in terms of a potential comparison with the phase
structure of QCD. However, it is important to keep in mind that much
of the interesting physics in QCD at finite density -- see, \eg
\cite{qcd} -- is associated to the fact that baryon number in QCD is
only carried by fermionic fields (quarks). This leads to the
existence of a Fermi surface at finite chemical potential. In gauge
theories dual to Dp/Dq systems as those considered here, baryon
number is also carried by scalar fields, and so the physics at
finite chemical potential is likely to be very different. In
particular, a chemical potential for charged scalars acts
effectively as a negative mass squared. In the case of free massless
scalars this leads to an instability. The theories considered here,
however, contain interaction, quartic terms in the fundamental
scalars, and so the chemical potential will presumably lead to
condensation of the scalars if these are sufficiently light.

\acknowledgments It is a pleasure to thank S.~Hartnoll,
T.~Hiramatsu, Y.~Hisamatsu, G.~Horowitz, A.~Karch, P.~Kumar, A.~Naqvi,
Y.~Sendouda, D.T.~Son and A.O.~Starinets for useful conversations.
Research at the Perimeter Institute is supported in part by funds
from NSERC of Canada and MEDT of Ontario. We also acknowledge
support from NSF grant PHY-0244764 (DM), NSERC Discovery grant
(RCM), JSPS Postdoctoral Fellowship for Research Abroad (SK), JSPS
Research Fellowships for Young Scientists (SM) and NSERC Canada
Graduate Scholarship (RMT). Research at the KITP was supported in
part by the NSF under Grant No. PHY99-07949.

While this paper was being prepared, we became aware of \cite{korea}
which overlaps considerably with the present work. We thank Sang-Jin
Sin for informing us of their project before publication.
Unfortunately we find no evidence of two phase transitions as
reported in \cite{korea}. We believe their result is spurious,
arising from including unphysical Minkowski embeddings in their
analysis.

\appendix

\section{Holographic dictionary}\label{holo}

As described in section \ref{Geom}, the D7-brane embeddings are
characterized by two nontrivial functions, $\chi(\rho)$ and
$A_t(\rho)$. Further, as usual in AdS/CFT-like dualities, the
asymptotic behaviour of these fields has a direct translation in
terms of operators in the dual gauge theory \cite{bigRev}. In
particular, considering the asymptotic behaviour in
eqs.~\reef{asymptA_t} and \reef{asymptD7}, the leading term
corresponds to the non-normalizable mode and its amplitude indicates
the coefficient with which the operator is added to the microscopic
Lagrangian of the field theory. Similarly, the subleading term is
the normalizable mode and its amplitude is proportional to the
vacuum expectation value of the operator. In the present case, since
we are discussing worldvolume fields, the corresponding operators
involve fundamental hypermultiplet fields.

Let us remind the reader that a hypermultiplet consists of two Weyl
fermions $\psi,\tilde{\psi}$ and two complex scalars $q,\tilde{q}$
-- the quarks of our theory. These are naturally organized so that
$\psi$ and $q$ transform in the fundamental of the $SU(\nc)$ gauge
group, while $\tilde{\psi}$ and $\tilde{q}$ transform in the
antifundamental. Further, with $\nf$ flavours (of equal mass), the
hypermultiplets transform under a global $U(\nf) \simeq SU(\nf)
\times U(1)_\mt{q}$ symmetry. The charges of the fields under the
diagonal $U(1)_\mt{q}$ are +1 for $\psi$ and $q$ and --1 for
$\tilde{\psi}$ and $\tilde{q}$. Hence the $U(1)_\mt{q}$ charge
naturally counts the net number of quarks in a given state. As the
colour group is $SU(\nc)$, baryons are composed of $\nc$ quarks and
so we would divide by $\nc$ for the number of baryons.

Now the operators dual to $\chi(\rho)$ and $A_t(\rho)$ can be
determined by considering the interactions of the open strings on
the D3/D7 array \reef{D3D7} before the decoupling limit
\cite{joebook}, in analogy with the closed strings. Such an
exercise leads to the following two operators:
\beqa A_t&\leftrightarrow&{\cal O}_\mt{q}=
{\psi^\dagger}\psi+\tilde{\psi}\tilde{\psi}^\dagger
+i\left({q}^\dagger{\cal D}_t {q}-({\cal D}_t{q})^\dagger
{q}\right)+ i\left(\tilde{q}\,({\cal D}_t \tilde{q})^\dagger-{\cal
D}_t\tilde{q}\,\tilde{q}^\dagger\right) \,,\nonumber
\labell{charge}\\
%
\chi &\leftrightarrow&{\cal O}_\mt{m}=
i\tilde{\psi}\psi+\tilde{q}(\mq+\sqrt{2}\Phi) \tilde{q}^\dagger +
{q}^\dagger(\mq+\sqrt{2}\Phi) {q} + h.c.\,.\labell{mass}
 \eeqa
Recall that the global flavour symmetry discussed above is the
$U(\nf)$ gauge symmetry of the $\nf$ D7-brane worldvolume. Hence
${\cal O}_\mt{q}$ is simply the quark charge density, \ie the time
component of the conserved $U(1)_\mt{q}$ current gauged by $A_\mu$
on the D7-brane. Note that the ${\cal D}_t$ indicate covariant time
derivatives in the $SU(\nc)$ gauge theory. The operator ${\cal
O}_\mt{m}$ is the variation of the mass term in the microscopic
Lagrangian, \ie ${\cal O}_\mt{m}=-\partial_{\mq}{\cal L}$.
Note that $\Phi$, one of the adjoint scalars in the ${\cal N}=4$
supermultiplet, as well as $M_q$, appear in the scalar terms here
after solving for the auxiliary field constraints within the full
coupled theory. As a check, one can observe that both of these
operators have conformal dimension\footnote{This dimension applies
in the UV where the effects of quark mass are negligible and the
theory becomes conformal.} $\Delta=3$, which matches the standard
prescription for the asymptotic powers appearing in
eqs.~\reef{asymptA_t} and \reef{asymptD7}.

Now we can make the dictionary between the asymptotic coefficients
and the dual gauge theory parameters precise by realizing that the
hypermultiplet states are the ground states of the 3-7 and 7-3
strings. Hence in the decoupling limit, these become precisely
strings stretching between the D7-brane and the horizon of the
D3-brane. For example, the quark mass is trivially derived for the
brane array \reef{D3D7} in asymptotically flat space. As this brane
configuration is supersymmetric at $T =0$, this mass persists in the
decoupling limit, where it is again the energy of a string
stretching between the D3- and D7-branes. This gives a relation
between $\mq$ and the parameter $m$ appearing in
eq.~\reef{asymptD7}. Further this relation is inherited by the
theory at finite temperature, since setting $T \neq 0$ does not
alter the asymptotic properties that determine the gauge theory
parameters.\footnote{Note that here we are referring to the bare or
current quark mass. The constituent quark mass is certainly modified
by thermal effects, as calculated in \cite{long,seat}.} Using this
result, we can formulate a variational argument \cite{toward} to
relate the second coefficient in the asymptotic expanision of $\chi$
to the field theory condensate
$\langle i(\tilde{\psi}\psi -\psi^\dagger\tilde{\psi}^\dagger)
\rangle$.\footnote{Ref.~\cite{toward} argued that the scalars would
not contribute to the condensate and in writing eq.~\reef{mc}, we
have certainly ignored such scalar terms. However, strong coupling
infrared dynamics might generate a condensate for the scalars. See
\cite{long} for an interesting example where the scalars dominate
this expectation value at high temperatures.} As the details of this
analysis can be found elsewhere \cite{toward,long},\footnote{Note
that the factor of $\nf$ in the formulae for $\qc$ was overlooked in
\cite{prl}. Further \cite{toward} only considers the case $\nf=1$. A
full analysis appears in \cite{long}.} we simply present the
results:
\beqa \mq &= \frac{u_0}{2^{3/2} \pi \ls^2}\, m \qquad \qquad \qc&= -
2^{3/2}
\pi^3 \ls^2 \nf T_\mt{D7} u_0^3\, c \nonumber\\
&=\frac{1}{2}\sqrt{\lambda}\,T\,m
\qquad\qquad\quad\hfill\hfill&=-\frac{1}{8}\sqrt{\lambda}\,\nf\,\nc
T^3\, c\, . \labell{mc} \eeqa
Given this result, we note that various figures were plotted in
terms of $T/\mbar\equiv 1/m$ and hence the relevant mass scale in
these plots is $\mbar=2 M_q/\sqrt{\lam}$. Up to a factor of $2\pi$,
this corresponds to the mass gap in the meson spectrum (at $d=0$)
\cite{us-meson,hmess}.

Now let us turn to the relation between the D7-brane gauge field and
the quark-charge operator \reef{charge}. Here the asymptotic value
of the potential $A_t(\infty)$ is proportional to the coefficient
with which the charge density ${\cal O}_\mt{q}$ enters  the microscopic
Lagrangian. This operator is normalized so that acting on a particular
state it yields exactly the net quark density, and therefore
the corresponding coefficient is precisely the chemical
potential $\mu$ for the quarks. Similarly the relevant expectation
value is the quark density $n_\mt{q}=\langle{\cal
O}_\mt{q}\rangle$.

Next we provide a precise definition of the particle density on the
string side of the duality. First, recall that the electric field on
the worldvolume can be thought of as arising from fundamental
strings `dissolved' into the the D7-brane \cite{bound}. The density
of these strings can be determined from the local charge density for the
two-form $B$-field. The standard convention is that the fundamental
string couples to the NS two-form through the worldsheet interaction
$T_\mt{f} \int d^2 \sigma\, B$. Hence a string pointing along the
$x^i$-axis sources $B_{ti}$ with the charge being just the string
tension $T_\mt{f}=1/(2 \pi \ls^2)$. Further, the one-form gauge
invariance of $B$ requires that the D7-brane action only involves
the combination $B+2\pi\ls^2\,F$. Hence we have
\beq \frac{\delta \ids}{\delta B_{ti}}=\frac{1}{2\pi
\ls^2}\frac{\delta \ids}{\delta F_{ti}}\, . \labell{equiv} \eeq
Combining these observations, we first conclude that since the
D7-brane carries an electric field in the $\trho$ direction, the
worldvolume effectively contains strings stretching along the radial
direction with a density precisely determined by the electric
displacement $d=-\delta \ids/\delta F_{t\trho}$. The minus sign in
the last expression means that for positive $d$ the strings are
oriented to be inward pointing towards the horizon at $\trho=1$.
Since the number of strings corresponds precisely to the number of
quarks in the field theory, the density of quarks is given by
integrating the string density on the D7-branes over the internal
three-sphere:
\beq \nq = \int d\Omega_3\, d = 2\pi^2\, d \, .\labell{ad} \eeq
While $d$ is not precisely the coefficient of the normalizable mode
in eq.~\reef{asymptA_t}, the two satisfy the simple relation given
in eq.~\reef{d}.

As noted above, the non-normalizable mode $A_t(\infty)$ indicates
that the charge density operator ${\cal O}_\mt{q}$ enters the
microscopic Lagrangian. As we wish to relate this bulk mode to the
chemical potential $\mu$ for the quarks in the microscopic theory,
it is natural to frame the discussion in terms of the grand
canonical ensemble. There the chemical potential enters the
partition function as
\be \exp\left[-\beta\int d^3x\,W(\beta,\mu)\right]\equiv
\sum\,\exp\left[-\beta\int d^3x\,\left({\cal H}- \mu\, {\cal
O}_\mt{q}\right) \right]\labell{parti}\ee
where a sum over all states is denoted on the right hand side. Of
course, $W(\beta,\mu)$ and ${\cal H}$ are the Gibbs free energy and
Hamiltonian densities, respectively. We know that $\mu\propto
A_t(\infty)$ but we would like to determine the exact constant of
proportionality. Towards this end, we note that, as can be seen from
eq.~\reef{parti},
%
\beq \frac{\delta W}{\delta \mu} = -\langle\, {\cal O}_\mt{q}\,
\rangle = -\,\nq \,. \labell{wow2}\eeq
To compare to the string description, we turn to the semiclassical
analysis of the Euclidean supergravity path integral, as described
in section \ref{thermal}. The grand canonical ensemble is
represented by the usual path integral with fixed $A_t(\infty)$ and
the on-shell action gives the leading contribution to the Gibbs free
energy, \ie $\ide=\beta\,W$. Hence to compare to eq.~\reef{wow2}, we
need to evaluate the change of the on-shell D7-brane action induced
by a change of the boundary value $A_t(\infty)$. Given the
worldvolume action (\ref{action}), the desired variation is
%
\beq \delta W =  \int d\trho\, d\Omega_{3} \, \delta {\cal L}_\mt{E}
= 2\pi^2\int_1^\infty d\trho\,  \frac{\delta{\cal L}_\mt{E}}{\delta
\pa_\trho A_t}\, \pa_\trho\delta A_t \,, \labell{vary}\eeq
where $\cal L$ is the D7-brane Lagrangian density. In
eq.~\reef{vary}, we have only integrated over the internal
three-sphere and the radial direction to produce the free energy
density in the gauge theory directions. Once again we recognize the
first factor as $d=\delta{\cal L}/\delta \pa_\trho A_t=-\delta{\cal
L}_\mt{E}/\delta \pa_\trho A_t$ (in the current notation -- note
that we need to distinguish between the `Lagrangian' densities
appearing in the Minkowski \reef{action} and Euclidean
\reef{dimensionless} actions), which is a constant on-shell. Hence
eq.~\reef{vary} reduces to
%
\beq \delta W = -2\pi^2\,d\,\left( \delta A_t(\infty)-\delta
A_t(1)\right) =-\,\nq\,\delta A_t(\infty)\, \labell{wow}\eeq
where we used \eqn{ad} and the fact that $A_t$ always vanishes on
the horizon so that we must have $\delta A_t(1)=0$. Finally
comparing eqs.~\reef{wow2} and \reef{wow}, we find
\beq A_t(\infty)=\mu\,, \label{wow3} \eeq
and so, as anticipated in the main text, the constant part of the
asymptotic gauge potential is precisely the chemical potential for
the quarks. If we wish to express results in terms of a baryon
chemical potential, we would convert $\mu_\mt{b}=\nc\,\mu$.

Let us also recall the formulae for the dimensionless quantities
defined in eqs.~\reef{redef} and \reef{connect} and which appear in
our calculations:
\beqa \tilde{\mu} &=& \frac{2\pi \ls^2 \mu
}{u_0}=\sqrt{\frac{2}{\lam}}\,\frac{\mu}{T}\,,\labell{for1}\\
\tilde{d} &=& \frac{d}{2 \pi \ls^2 u_0^3 \nf T_\mt{D7}}=
\frac{2^{5/2}}{\nf\nc\lam^{1/2}}\,\frac{\nq}{T^3}
=\frac{2^{5/2}}{\nf\lam^{1/2}}\,\frac{\nb}{T^3}\,.\labell{for2}
 \eeqa
Hence as with the previous definitions, the temperature $T$ provides
the scale to make these quantities dimensionless but implicitly we
have also introduced interesting factors of the 't Hooft coupling,
as well as of $\nf$ and $\nc$. In particular, we see that
$\tilde{d}$ is naturally related to the expectation value of the
baryon number $\nb$ in \reef{for2}.

\end{document}